\documentclass[preprint,12pt]{elsarticle}

\usepackage{amssymb}
\usepackage{xcolor}
\usepackage{tabu}
\usepackage{amsmath}
\usepackage{multirow}
\usepackage{multicol}
\usepackage{csquotes}
\usepackage{url}

\journal{Computers \& Security}

\begin{document}

\begin{frontmatter}


\title{A Comprehensive Analysis of Adversarial Attacks against Spam Filters}

\author[inst1]{Esra Hotoğlu}

\affiliation[inst1]{organization={WISE Lab., Department of Computer Engineering},
            addressline={Hacettepe University}, 
            city={Ankara},
            country={Turkey}}

\author[inst1]{Sevil Sen\corref{cor1}}
\ead{ssen@cs.hacettepe.edu.tr}
\cortext[cor1]{Corresponding author}

\author[inst2]{Burcu Can}

\affiliation[inst2]{organization={Department of Computing Science and Mathematics},
            addressline={University of Stirling}, 
            city={Stirling},
            country={UK}}

\begin{abstract}
Deep learning has revolutionized email filtering, which is critical to protect users from cyber threats such as spam, malware, and phishing. However, the increasing sophistication of adversarial attacks poses a significant challenge to the effectiveness of these filters. This study investigates the impact of adversarial attacks on deep learning-based spam detection systems using real-world datasets. Six prominent deep learning models are evaluated on these datasets, analyzing attacks at the word, character sentence, and AI-generated paragraph-levels. Novel scoring functions, including spam weights and attention weights, are introduced to improve attack effectiveness. This comprehensive analysis sheds light on the vulnerabilities of spam filters and contributes to efforts to improve their security against evolving adversarial threats. 
\end{abstract}



\begin{keyword}
email security \sep spam detection \sep adversarial learning \sep natural language processing \sep deep learning
\end{keyword}

\end{frontmatter}


\section{Introduction}
Deep learning has seen significant advancements in the field of natural language processing (NLP), particularly in tasks such as email filtering. Email filters play a critical role in detecting spam, viruses, and malware, serving as the first line of defence against cyber-attacks. Cybercriminals often target personal and valuable data, such as cryptocurrency wallets and email credentials, so robust email filtering is essential to protect users from potential security breaches. 

According to Cybersecurity Report of Trend Micro \cite{r4}, the increase in malware detections and Business Email Compromises (BECs) from 2022 to 2023 indicates increasingly sophisticated methods. In addition to subtle tactics to trick users into clicking on malicious links, spam campaigns remain effective and can bypass email filters. In addition, the FBI's 2023 Internet Crime Report \cite{r5} indicates a significant increase in the frequency and financial impact of online fraud. Phishing scams, in which cybercriminals impersonate legitimate companies to obtain personal and financial data via email, were the most common type of reported fraud. Business Email Compromise (BEC) has been identified as one of the most expensive types of fraud, with 21,489 complaints resulting in \$2.9 billion in losses. 

Google, Outlook and Yahoo use different methods for spam filtering to filter out unwanted messages. Google Mail (Gmail) classifies emails as spam, promotional, or social based on their content. Google's data centers use hundreds of rules to determine whether an email is valid or spam. Outlook, on the other hand, automatically filters spam, and users can easily create custom rules to further categorize emails. The Yahoo email provider also has its own algorithms in order to detect spams \cite{r1}. Gmail, used by millions, has advanced security features to block 99.9\% of spam, phishing, and malware, and uses TensorFlow to improve spam email detection capabilities \cite{r2}. In addition, Yahoo filters are reported to be 99.9\% successful at catching spam, malware and phishing emails \cite{r3}.

Despite their effectiveness, email spam filters can be manipulated, particularly through adversarial learning techniques. Adversarial learning, a prominent method in machine learning, involves deliberately introducing small changes to the input data to fool a model, causing it to misclassify or make incorrect predictions. This phenomenon has become a significant problem, particularly in the field of deep learning, where even state-of-the-art classifiers can be vulnerable to such attacks. Adversarial attacks on machine learning models typically fall into two broad categories: white-box attacks and black-box attacks. In white-box attacks, the adversary has complete access to the model, including its structure, parameters, and training data. In contrast, in black-box attacks, the adversary has limited or no access to the inner workings of the model, relying instead on external observations to construct adversarial inputs.

Moreover, AI-generated emails pose a significant threat to email spam filters. Through the use of advanced deep learning algorithms and natural language processing (NLP), AI can create content that closely mimics human writing. This means that AI-generated spam emails can appear highly convincing and may bypass traditional spam filters designed to catch more obvious threats. As a result, malicious actors can exploit this technology to produce sophisticated spam messages that deceive recipients. These deceptive emails can manipulate into revealing sensitive information, clicking on harmful links, or engaging in other actions that compromise their security. This evolving challenge underscores the need for more advanced and adaptive security measures to detect and mitigate AI-driven threats.

Recently, significant efforts have been made to develop deep learning-based systems, primarily utilized for natural language processing tasks, given the discrete nature of text. Nevertheless, adapting similar attacks to the NLP domain has proven challenging due to this inherent characteristic. Therefore, there is a growing body of research that focuses on adversarial examples in text-based systems. This study is one of them by putting emphasis on spam filters. It investigates the impact of deliberate perturbations of input vectors on various advanced spam filters using three prominent real-world text datasets commonly used in spam email research: SpamAssassin \cite{d1}, Enron Spam \cite{d2}, and TREC 2007 \cite{d3}. It thoroughly analyzes the generation of black-box attacks that target spam filters at multiple levels, including character, word, and sentence levels. 

These attacks are designed to generate adversarial examples that are capable of bypassing various spam detection filters. These filters are based on a variety of deep learning architectures tailored to various tasks and data structures: a Long Short Term Memory (LSTM) model for sequential data, a Convolutional Neural Networks (CNN) model for spatial features, a Feed Forward Neural Network with Dense layers for general tasks, an attention model for selective focus, and a transformer model for efficient sequence processing, and distilBERT model which is a pre-trained model for efficient and compact language understanding. In addition, novel scoring functions are introduced to generate more effective adversarial attacks. Performance evaluation of the proposed scoring functions involves subjecting them to rigorous testing against various black-box attack scenarios and comparison with existing scoring methods used in spam filtering systems. Additionally, AI-generated paragraph-level attack that includes spam and non-spam emails are also tested on these filters to assess their effectiveness. Specifically, it examines how these AI-generated emails interact with and potentially bypass existing spam filters.

This comprehensive analysis aims to contribute to ongoing efforts to improve the security and resilience of spam detection filters in response to evolving adversarial threats. In summary, this study entails analyzing a range of adversarial attacks designed to undermine spam email detection systems. The primary contributions of the study are highlighted as follows: 
\begin{itemize}
    \item Six prominent deep learning-based spam detection systems are developed and thoroughly evaluated against adversarial attacks using three real-world datasets. Unlike many studies in the literature \cite{b1, b3, b4, b5, b6, b7, b8, b9, b10, b11}, which often limit their evaluations to a single dataset, our study provides a more comprehensive assessment. Furthermore, most studies in the literature focus on traditional models and typically only examine one or two deep learning algorithms. This study tests the six prominent deep learning-based models against adversarial attacks. 
    \item Adversarial attacks against spam filters are comprehensively analyzed at four levels: word-level, character-level, sentence-level, and AI-generated paragraph-level. While previous studies 
    have predominantly concentrated on word-level attacks only, with only a single study \cite{b1} addressing into character-level attacks, our research addresses all potential attacks at each level. Sentence-level attacks are investigated for the first time against NLP-based systems in this study. Therefore, this comprehensive analysis ensures a thorough examination of the effectiveness and vulnerabilities of spam filters, leading to a more robust understanding of their resilience in real-world scenarios.
    \item This study introduces novel scoring functions, namely spam weights and attention weights scoring functions to identify the most effective words in order to create more effective attacks in the field of spam detection. Their effectiveness are demonstrated in the results. 
    \item This study also investigates the impact of AI-generated paragraph-level spam and non-spam emails on spam detection systems. This investigation provides insights into the challenges faced by spam detection technologies and helps identify potential areas for improvement.
\end{itemize}

The paper is organized as follows: Section 2 provides a literature review on attacks implemented in spam filters. Section 3 outlines the datasets used in the study, details the preprocessing steps, and describes the deep learning models used for spam detection. It also introduces the adversarial attacks and the associated scoring functions. Sections 4 and 5 present and discuss the experimental results. Finally, Section 6 provides concluding remarks on the work.

\section{Related Work}
Email spam detection is crucial for protecting users from unwanted messages, phishing attempts, malware distribution, and other security threats. It involves analyzing incoming email messages to distinguish between ham (non-spam) and spam content. Various algorithms and techniques are commonly employed for spam detection. Unlike traditional classifiers, deep learning models offer the ability to learn abstract features. Deep learning techniques, such as Long Short-Term Memory Networks (LSTMs), Convolutional Neural Networks (CNNs), attention mechanisms, and transformer architectures, are particularly effective for feature extraction and classification in spam detection tasks, especially when dealing with complex data such as images or large text corpora. These algorithms are often combined with feature engineering techniques, pre-processing steps, and evaluation metrics to construct robust and efficient spam detection systems.

Adversarial attacks are techniques used to deceive or manipulate machine learning models through the input of carefully crafted data. These attacks target weaknesses in the model's decision-making processes, often leading to misclassifications or other unwanted outcomes. Adversarial attacks can manifest in various ways, such as by adding barely detectable noise to input data, altering pixels in images, or changing features in text.

In the realm of adversarial attacks two primary strategies stand out: white-box attacks and black-box attacks. In a white-box attack scenario, the adversary has full insight into the target model, including its structure, parameters, loss functions, activation functions, as well as access to both input and output data. This level of access enables the attacker to meticulously craft adversarial perturbations tailored to exploit vulnerabilities in the model. By approximating the worst-case scenario for a given model and input, white-box attacks pose a significant threat, often achieving high success rates in compromising model integrity and performance. This adversary strategy is particularly potent in controlled environments where the attacker has unrestricted access to the model's inner workings \cite{s1, s3}.

Conversely, black-box attacks operate under the assumption that the attacker lacks detailed knowledge, such as its architecture and parameters. However, black-box attackers still have access to the model's input and output interfaces, allowing them to query the model and observe its responses. In this scenario, attackers often rely on heuristic methods to generate adversarial examples, leveraging insights gained from probing the model's behavior through input-output interactions. In real-world scenarios, black-box attacks are often the most feasible and realistic approach. Despite the inherent limitations imposed by the lack of model transparency, black-box attacks remain a viable threat vector, highlighting the importance of developing robust defense mechanisms against adversarial manipulation \cite{s1, s3}.

While the general classifications of attacks provide a foundational framework, it's essential to recognize that for Natural Language Processing (NLP) tasks, attack strategies and types differ due to the unique characteristics of text data compared to image or audio data. Textual content presents distinct challenges and opportunities for adversarial manipulation, leading to specialized classifications of attack techniques tailored to NLP domains. In the context of NLP, attacks can be classified according to the level of granularity of modifications made to the text data. Specifically, three primary types of attack techniques emerge: character-level attacks, word-level attacks and sentence-level attacks. Each type targets different linguistic components within the text, allowing adversaries to exploit vulnerabilities in NLP systems effectively \cite{s5, s6}.

Character-level attacks involve the manipulation of individual characters within the text, such as inserting, removing, substituting, or rearranging characters to induce misclassification or alter semantic meaning. These attacks often capitalize on the subtle nuances of language to evade detection and compromise model integrity. Word-level attacks operate at the level of words, where adversaries modify or replace entire words within the text to deceive NLP models. By strategically choosing words or phrases, attackers can distort the intended message or inject malicious content without significantly altering the overall structure of the text. Sentence-level attacks focus on the manipulation of entire sentences or segments of text to influence model predictions or behavior. Adversaries may introduce grammatical errors, syntactic anomalies, or semantic inconsistencies to disrupt model performance or mislead downstream processing \cite{s5, s6}.

In recent years, the exploration of deep learning algorithms for spam detection has gained attention in the field of adversarial learning. As a result, there has been a significant amount of research on spam detection using adversarial machine learning. However, previous studies have primarily focused on the good word attack, which modifies spam emails by inserting or appending words that indicate a legitimate email. 

In \cite{b6}, a counter-attack strategy using multiple instance learning are proposed to defend good word attacks on statistical email spam filters. This study demonstrates that multiple-instance learners outperform standard single-instance learners, including logistic regression, support vector machine, and the commonly used Naive Bayes model, in withstanding good word attacks. Jorgensen et al. \cite{b7}, a similar multiple instance learning counter-attack strategy is presented to combat adversarial good word attacks on statistical spam filters. This involves transforming each email into a collection of multiple segments and applying multiple sample logistic regression to these collections. The introduced classifier is claimed to be more robust against good word attacks compared to commonly used methods in the spam filtering domain. 

Furthermore, in \cite{b8}, the performance of Naive Bayes and maximum entropy spam filters is examined in response to active and passive good word attacks. The study determines the effectiveness of a word by averaging the weights of all the words in each filter. The results suggest that adding a relatively small number of easily identifiable words can allow around 50\% of currently blocked spam to pass through a spam filter. Another study \cite{b9} highlights the ease of implementing some attacks and their varying effectiveness, noting that while some methods like the common word attack can be more efficient than others, they often only succeed against specific filters. It suggests that future efforts should include examining different spam evasion techniques, understanding vulnerabilities in various filters, and exploring the impact of retraining filters.

A novel attack method is proposed in \cite{b2}, involving the alteration of textual data by using NLP based on the results of constructed adversarial samples designed to deliberately modify the features representing an email. Various natural language feature extraction approaches, such as TF-IDF, Word2vec, and Doc2vec, are compared against white-box attacks. By conducting experiments on various datasets and utilizing various classification models such as Support Vector Machine (SVM), decision tree, logistic regression, Multi-layer Perceptron (MLP), and ensemble classifiers. The proposed method is demonstrated to be capable of crafting adversarial examples in the text domain, significantly degrading the accuracy of spam detection systems. In \cite{b3}, researchers explore the impact of adversarial scenarios on machine learning-based methods such as email spam filters. Three invasive techniques are tested using NLP along with a Bayesian model: synonym replacement, raw word injection, and spam word spacing, demonstrating their effectiveness in deceiving machine learning models.

In addition, Ozkan et al. \cite{ozkan2019analysis} investigates how adversarial attacks affect conventional spam detection systems that use machine learning models such as Naïve Bayes (NB) and Support Vector Machines (SVM). Four types of attacks, tokenization, obfuscation, word addition, and word substitution, were tested to evaluate their effects on spam filter accuracy. Results show that while tokenization and obfuscation have limited effects, word addition and word substitution attacks significantly reduce filter accuracy, potentially rendering the filters ineffective. Also, in \cite{b4}, two innovative text generation methods are introduced to enhance the effectiveness of attacks by leveraging adversarial perturbations produced through adversarial example generation algorithms. One method approximates TF-IDF values in the adversarial examples, while the other incorporates special words into the original emails. The study employs the Projected Gradient Descent (PGD) algorithm and evaluates its performance across various machine learning classification models, including SVM, K-Nearest Neighbors (KNN), decision trees (DT), and logistic regression (LR), under both white-box and black-box attack scenarios. In another study \cite{b5}, a defense mechanism is proposed to mitigate the impact of these ideal poisoning attacks on linear classifiers, based on outlier detection. However, since the attack strategies do not consider detectability constraints, the resulting counterexamples are notably different from real data points. The findings indicate that less aggressive attacks, like label flipping, can be challenging to detect with these defense mechanisms, as the generated attack points closely resemble real data points.

Moreover, Nelson et al. \cite{b10} illustrate how the SpamBayes spam filter can be effectively neutralized with minimal knowledge of the system and restricted access to the training data. While they present successful defenses such as the RONI defense, that blocks messages from dictionary attacks completely, and the dynamic threshold defense, which mitigates the impact of dictionary attacks, they highlight the persistent challenge of defending against focused attacks due to the attacker's additional knowledge. On the other hand, Gu et al. \cite{b11} proposed marginal attack methods to deceive a Naive Bayesian spam filter by adding sensitive words to sentences. Three strategies for selecting sensitive words are proposed, resulting in significant reductions in the filter's detection accuracy. These attacks significantly reduce the filter's accuracy, even with just one word added. The study also showed that the generated adversarial examples can disrupt other traditional filters such as logistic regression, decision tree, and linear support vector machine.

The previous studies have mainly focused on word-level attacks, but there are also a few studies investigating the character-level attacks, also using deep learning algorithms. For instance, in \cite{b1}, a new algorithm named DeepWordBug is introduced. This algorithm efficiently generates minor text perturbations at character-level within a black-box environment, compelling deep learning classifiers to misclassify text inputs. Their evaluation is carried out on two real text datasets containing Enron spam emails and IMDB movie reviews, and includes the development of scoring strategies to identify the most critical words for modification, leading to incorrect predictions. Remarkably, their results illustrate a significant reduction in classification accuracy, decreasing from 99\% to 40\% on the Enron Spam dataset and from 87\% to 26\% on the IMDB dataset. Furthermore, Boucher et al. \cite{s2} analyzes a broad range of adversarial examples across various domains, beyond spam filters, capable of attacking text-based models at the character level in a black-box setting. They employ perturbations to manipulate the output of various NLP-based systems. The study demonstrates that attacks involving invisible characters, homoglyphs, reordering, or deletion could substantially impair the performance of vulnerable models. 

Zhang et al. \cite{s1} present the first comprehensive research on generating textual adversarial examples on deep neural networks. They reviewed recent research efforts and research studies that produced textual adversarial examples on DNNs. They also comprehensively collected, summarized and analyzed these studies and ensured that the article was self-contained by covering all relevant information. Finally, they have provided an excellent resource for researchers to understand the challenges, techniques, and key topics in this field. In another research \cite{s3}, various forms of adversarial attacks on machine learning in the context of network security are examined and two novel classification frameworks are introduced for detecting and mitigating such attacks. First, the attacks are classified based on the classification of network security applications. Then, they are classified according to the problem domain and classification model. Finally, an in-depth analysis of diverse defense strategies aimed at protecting machine learning-based network security applications from adversarial attacks are analyzed. 

\begin{table*}[!t]
\centering
\caption{Analysis of Previous Studies }
\resizebox{\textwidth}{!}{\begin{tabu}{l l l l l}
\tabucline[2pt]{-}
\textbf{Previous Studies} & \textbf{Dataset} & \textbf{Methodology} & \textbf{Scoring Functions} & \textbf{Attacks}\\
\tabucline[2pt]{-}
Zhou et al. \cite{b6} & TREC 2006 &	Naive Bayes & - & Good Word Attack \\
\hline
Jorgensen et al. \cite{b7} & TREC 2006 & LR, Naive Bayes, SVM & - & Good Word Attack  \\
\hline
Lowd and Meek \cite{b8} & Hotmail Feedback Loop &  Naive Bayes, Maxent & - & Good Word Attack  \\
\hline	
Wittel et al. \cite{b9} & SpamAssassin & SpamBayes & - & \shortstack[l]{Dictionary Word Attack  \\ Common Word Attack} \\ 
\hline 	
Cheng et al. \cite{b2} & Ling, Tutorial, Enron Spam & SVM, DT, LR, MLP & - & PGD attack \\
\hline
Kuchipudi et al. \cite{b3} & SMS Spam & Naive Bayes & - & \shortstack[l]{Synonym Replacement \\ Ham Word Injection \\ Spam Word Spacing} \\
\hline
Chenranc et al. \cite{b4} & Enron Spam  & SVM, KNN, DT, LR & - & \shortstack[l]{PGD Attack, \\ Adding Special Words}  \\ 
\hline
Paudice et al. \cite{b5} & Spambase  & Linear Classifier & - & Poisoning Attacks \\	
\hline 
Nelson et al. \cite{b10} & TREC 2005 & SpamBayes & - & \shortstack[l]{Dictionary Attack  \\ Focused Attack}\\ 	
\hline
Ozkan et al. \cite{ozkan2019analysis} & SpamAssassin, Enron Spam & SVM, NB & - & \shortstack[l]{Tokenization, Obfuscation \\ Word Addition \\ Word Substitution} \\ 	
\hline
Gu et al. \cite{b11} & SMS Spam & Naive Bayes, SVM, DT, LR & - & Word Addition \\ 	
\hline
Gao et al. \cite{b1} & Enron Spam & LSTM, CNN & \shortstack[l]{Replace-1 Score, \\ Temporal Head Score, \\ Temporal Tail Score, \\ Combined Score} & \shortstack[l]{Substitution, Deletion Chars \\ Insertion, Swap Chars}  \\ 	
\hline
Our Study & \shortstack[l]{SpamAssassin, \\ Enron Spam \\ TREC 2007} & \shortstack[l]{LSTM, CNN \\ Dense, Attention \\ Transformer} & \shortstack[l]{Replace-1 Score, \\ Spam Weights, \\ Attention Weights} & \shortstack[l]{Out of Vocab, Deleting Words \\ Synonym Replacement \\ Antonym Replacement \\ Insertion \& Deletion Chars \\ Replacement \& Swapping Chars \\ Add Ham, Spam Sentence \\ Ham-Spam Sentences}  \\ 
\tabucline[2pt]{-}
\end{tabu}}
\label{tab_prestudies}
\end{table*}

On the other hand, AI-generated content has increasingly influenced deep learning models for spam detection in recent years. A study \cite{ai1} explores how Large Language Models (LLMs) like GPT-3.5, Bard, and BingAI generate datasets for password strength prediction. The research highlights the potential and limitations of LLMs for data creation and encourages further work to enhance their capabilities and data diversity. Also, artificial intelligence is widely used to produce images, as discussed in \cite{ai2}, which evaluates six AI-generated-image detection methods across 23 datasets, including images from GANs, diffusion models, and transformers, highlighting the widespread use of artificial intelligence in image generation. At the same time, artificial intelligence plays a crucial role in both generating and detecting spam emails, as discussed in \cite{ai3, ai4, ai5}. These sources examine various AI-based spam detection models, assess their performance on multiple datasets, and emphasize the growing importance of AI in enhancing email security and filtering systems.

The related studies on adversarial attacks against spam filters are summarized in Table \ref{tab_prestudies}. As shown, there are only a few studies focusing on adversarial attacks against spam filters that utilize deep learning algorithms, despite the prevalence of such algorithms in many modern spam filters. On the contrary, our study centers on the exploration of spam filters employing various deep learning techniques. Moreover, while previous studies have generally concentrated on word-level attacks only, our study comprehensively analyzes possible attacks at the character, word, and sentence levels. Additionally, by examining such attacks in black-box scenarios, we aim to simulate real-world scenarios more accurately. Last but not least, we propose different scoring functions to select words for these attacks, thereby enhancing their effectiveness. As these attacks play a crucial role in assessing the robustness of models against adversarial attacks, they can be integrated into the training of deep learning models to improve spam classifiers. To sum up, this study provides a comprehensive analysis of adversarial attacks against modern spam filters, filling a notable gap in existing research.

\section{Methodology}

This study targets the bypassing of several neural network architectures by adversarial attacks. These models include Long Short-Term Memory (LSTM) networks, a specialized version of Recurrent Neural Networks (RNN), Convolutional Neural Networks (CNN), a Feed Forward Neural Network with Dense layers, an LSTM model with a single attention layer, a transformer model and a pre-trained model called distilBERT. The primary objective is to illustrate the impact and extent of various attack types on various deep learning spam filters.

In black-box attacks, adversaries can only modify the test data without access to the filters. This study uses three well-known spam datasets to train spam filters and generate adversarial attacks. First, preprocessing, tokenization and sequencing steps are applied to all datasets. Subsequently, spam filters based on LSTM, CNN, LSTM with attention and the transformer are developed using the Keras and TensorFlow libraries. The distilBERT is utilized through Hugging Face's Transformers library. Finally, different types of adversarial attacks at different levels (character, word, sentence, and AI-generated paragraph-level) with different scoring functions are executed against these DL-based spam filters and a thorough evaluation is performed.

\subsection{Datasets and Preprocessing}
The three datasets used in this study, namely SpamAssassin \cite{d1}, Enron Spam \cite{d2} and TREC2007 \cite{d3}, are summarized below:
\begin{itemize}
\item {\verb|SpamAssassin|}: The dataset is obtained from the Apache Public Datasets and the Apache SpamAssassin Projects, which maintain a repository of archived emails. This dataset consists of 2,400 spam and 6,954 ham (i.e. not spam) emails \cite{d1}. 
\item {\verb|Enron Spam|}: The dataset is collected from the mailboxes of Enron employees, in the cleaned-up form provided, which includes only ham messages, and from four different sources for spam messages\cite{d22}. It contains 17,171 spam emails and 16,545 ham emails \cite{d2}.
\item {\verb|TREC2007|}: The TREC (Text Retrieval Conference) 2007 Public Corpus Dataset was collected through tasks aimed at classifying email messages, with variations in the amount and frequency of feedback received by the system. It contains 50,199 spam emails and 25,220 ham emails \cite{d3}. 
\end{itemize}

\begin{table}[htbp]
\caption{Distribution of Datasets}
\begin{center}
\small
\begin{tabular}{c c c c c}
\hline
 &  \multicolumn{2}{c}{\textbf{Spam Emails}} & \multicolumn{2}{c}{\textbf{Ham Emails}} \rule[1ex]{0pt}{1.5ex} \\
\hline
\textbf{Dataset} & \textbf{Train Set} & \textbf{Test Set} &\textbf{Train Set} & \textbf{Test Set} \rule[1ex]{0pt}{1.5ex} \\
\hline
SpamAssassin & 1920  & 480    & 5563  & 1391 \rule[1ex]{0pt}{1.5ex} \\
\hline
Enron Spam   & 13,737 & 3434  & 13,236 & 3309 \rule[1ex]{0pt}{1.5ex} \\
\hline
TREC2007     & 40,159 & 10,040 & 20,176  & 5044 \rule[1ex]{0pt}{1.5ex}  \\
\hline
\end{tabular}
\label{tab_dataset}
\end{center}
\end{table}

These corpora were chosen because of their widespread use in spam-related studies. Therefore, the use of these datasets will allow an easy comparison between our results and existing studies. 80\% of the data is used for training and 20\% for testing. The distribution of ham and spam in both training and testing datasets is given in Table \ref{tab_dataset}.

A number of preprocessing steps have been implemented to clean up the data and reduce the input size. Firstly, punctuation, numbers, hyperlinks, and stop words such as “the”, “a”, “an”, “in” are removed. Additionally, all text is converted to lowercase. Word stemming and lemmatization preprocessing techniques are also used. Both methods aim to simplify words to their basic forms. These preprocessing steps reduce the computational cost and have no negative impact on the classification results. 

Once the text is cleaned, it is converted into a numerical representation so that it can be used as input to the model. First, it is tokenized using the Keras tokenizer, which splits sentences into words and encodes them into integers. Next, each sentence is represented by sequences of numbers. Finally, padding is applied to ensure a uniform length for each sequence.

\subsection{Methods} 

Deep learning algorithms are used for spam detection because they offer many advantages over traditional methods for dealing with the complexities of data. Spam emails are becoming increasingly sophisticated, often imitating legitimate communications or using new techniques to evade detection. Spam detection requires understanding intricate patterns in email text, such as unusual phrasing, grammar, or subtle cues that indicate spam. Neural architectures such as Recurrent Neural Networks (RNNs) and Long Short-Term Memory (LSTM) networks are able to process sequential data, capturing long-term dependencies and patterns in sentences, which helps in identifying subtle and disguised spam content. This is crucial because spam messages often evolve to mimic regular email content. In addition, deep learning models are trained on large datasets and can handle diverse inputs, such as text, images, or hyperlinks within emails. Models like Convolutional Neural Networks (CNNs) can process image-based spam, and transformer-based models, such as BERT or GPT, can analyze the context of entire email bodies. This versatility allows spam filters to adapt to different formats of spam, whether it's text, attachments, or multimedia content.

On the other hand, transformer-based models such as BERT and GPT bring a deeper level of understanding by considering the context in which words occur and the cross-relationships between words, helping to learn deeper relationships beyond semantics. This is particularly important for spam detection because spammers often craft their messages to seem legitimate, using context-specific language. Deep learning models can distinguish subtle differences in how certain words are used, allowing them to recognize even cleverly disguised spam. Deep learning models can automatically learn relevant features from raw email data, reducing the need for manual feature engineering. CNNs, for instance, are excellent at extracting key features such as word patterns, that may indicate spam, while the attention mechanism can highlight important parts of an email that are more likely to indicate spam. This allows for a more efficient and accurate classification process. Moreover, emails contain a wide range of information, from subject lines to embedded links, multimedia content, and metadata. Deep learning models are capable of processing and analyzing meaningful insights, patterns or information from data with a large number of features and dimensions, efficiently. They prioritize important features while ignoring irrelevant ones, making the detection process more effective.

Hence, we employ six different classifiers based on the following deep learning architectures: Long Short-Term Memory Networks (LSTM), Convolutional Neural Networks (CNN), a fully connected neural network (dense network), an LSTM with an attention layer, a transformer, and distilBERT which is a lightweight, faster, and smaller version of the transformer-based BERT model in our analysis. While distilBERT is a pre-trained model, other classifiers are trained in this study.

Recurrent neural networks (RNNs) are able to capture sequential dependencies by incorporating loops into their structure. However, traditional RNNs faced challenges with backpropagation, which were addressed by Hochreiter and Schmidhuber \cite{m5} through the development of Long Short-Term Memory (LSTM) architectures. LSTMs have become one of the most favoured methods for text-based tasks. Many recent studies \cite{m1,m2,m3,m4} explore the effectiveness of LSTMs in various applications. Similarly, in the field of spam detection, studies \cite{m14,m15,m16} have used LSTMs and achieved high accuracies.

Convolutional Neural Networks (CNNs) are network architectures originally developed for image processing. They typically consist of convolution layers, pooling layers, and fully connected layers. Recent studies have demonstrated that CNNs are also effective for word-level text classification \cite{m6}. Several studies have used CNN filters to generate and evaluate adversarial text examples \cite{m1,m3,m4,m7,m8}. In addition, CNNs are widely used in spam detection and have shown promising results \cite{m16,m17,m18,m19}.

One of the latest advancements in deep learning is the integration of a mechanism known as attention \cite{m13}. This mechanism aims to identify the relationship between inputs and expected outputs, giving greater importance to relevant inputs. It has already been used for different tasks such as sentiment analysis \cite{m10, m11}. In attention mechanism, a context vector is shared between the input and the output. Attention weights indicate which words are useful for generating the desired output. The attention method, commonly used in the field of natural language processing, has found extensive application in spam detection studies \cite{m19,m20,m21,m22,m23}, providing robust approaches to detecting spam emails.

The transformer architecture, proposed by Vaswani et al. \cite{m24}, is an encoder-decoder model. This innovative design has gained popularity due to its parallelizability, scalability, and ability to capture long-term dependencies in sequential data without using recurrent connections as in RNNs. Comprising encoder and decoder components, the transformer architecture is structured around a self-attention mechanism that learns the importance of different parts of a sequence by attending to itself. This attention mechanism is run through several times in parallel, which is called multi-head attention. Its outstanding effectiveness in NLP tasks \cite{m25, m26} has established it as a cornerstone in the field. There have also been notable studies in spam detection \cite{m27, m28, m29}, where its ability to detect complex patterns in text data has been crucial in efficiently reducing spam emails.

A pre-trained model in machine learning, particularly in natural language processing (NLP) and computer vision, refers to a model that has already been trained on a large dataset and is subsequently used as a starting point for training on a specific task. Sanh et al. \cite{m30} demonstrated that smaller language models pre-trained with knowledge distillation can achieve similar performance on many downstream tasks. The distilBERT, an optimized version of the BERT (Bidirectional Encoder Representations from Transformers) model, is designed to be more compact and efficient while retaining much of BERT's performance. It is also utilized across a variety of natural language processing (NLP) tasks \cite{m34, m35}. Notable studies \cite{m31, m32, m33} on spam detection have shown that it is effective in providing high accuracy, improving performance, and optimizing resource utilization.

An attempt has been made to use systems similar to those examined in the previous studies to facilitate comparisons with them. Extensive testing was carried out before the final model parameters for each model were determined. This was accomplished by fine-tuning the models for the spam detection task using RandomizedSearchCV \cite{randomsearch}. It is a hyperparameter tuning technique in machine learning used to optimize the performance of a model, and it is part of the scikit-learn library. Instead of exhaustively searching over all possible combinations of hyperparameters (as in GridSearchCV), RandomizedSearchCV samples a fixed number of hyperparameter combinations from a specified distribution or range. This makes it more efficient, especially when dealing with a large number of hyperparameters. Parameters such as the number of input units for LSTM layers, the number of units for dense layers, number of filters, kernel size for convolutional layers, dropout rate, activation function, optimizer, learning rate, and loss function have been selected. As false positive rates have more serious implications than false negatives, a trade-off between lower false positive rates and accuracy values was considered. The architectural details of each model are shown in Table \ref{tab:model}. In the table, the "None" values in the Input Shape and Output Shape columns refer to dynamic or flexible dimensions in the model architecture. In deep learning models, the batch size is usually not specified when defining the model architecture, allowing the model to handle inputs of varying batch sizes during training or inference.

\begin{table}[!t]
\caption{Architectural Details of the Models}
\begin{center}
\scriptsize
\begin{tabular}{l l l l}
\hline\hline
\textbf{Model} & \multicolumn{3}{c}{\textbf{Layers}} \\
\hline\hline
& Layer Name & Input Shape & Output Shape \\
\hline\hline
\multirow{4}{*}{LSTM} 
& Input Layer & None, 350 & None, 350 \\
\cline{2-4}
& Embedding Layer & None, 350 & None, 350, 50 \\
\cline{2-4}
& LSTM Layer & None, 350, 50 & None, 32 \\
\cline{2-4}
& Dense Layer & None, 32 & None, 1 \\
\hline\hline
\multirow{7}{*}{Dense} 
& Input Layer & None, 350 & None, 350 \\
\cline{2-4}
& Embedding Layer & None, 350 & None, 350, 50 \\
\cline{2-4}
& Flatten Layer & None, 350, 50 & None, 17500 \\
\cline{2-4}
& Dense Layer & None, 17500	& None, 416 \\
\cline{2-4}
& Dropout Layer & None, 416 & None, 416 \\
\cline{2-4}
& Dense Layer & None, 416 & None, 416 \\
\cline{2-4}
& Dense Layer & None, 416 & None, 1 \\
\hline\hline
\multirow{7}{*}{CNN} 
& Input Layer & None, 350 & None, 350 \\
\cline{2-4}
& Embedding Layer & None, 350 & None, 350, 50 \\
\cline{2-4}
& Conv1D & None, 350, 50 & None, 348, 128 \\
\cline{2-4}
& GlobalMaxpooling1D & None, 348, 128 & None, 128 \\
\cline{2-4}
& Dense Layer & None, 128 & None, 192 \\
\cline{2-4}
& Dropout Layer & None, 192 & None, 192 \\
\cline{2-4}
& Dense Layer & None, 192 & None, 1 \\
\hline\hline
\multirow{5}{*}{Attention} 
& Input Layer & None, 350 & None, 350 \\
\cline{2-4}
& Embedding Layer & None, 350 & None, 350, 50 \\
\cline{2-4}
& LSTM Layer & None, 350, 50 & None, 350, 32 \\
\cline{2-4}
& Attention Layer & None, 350, 32 & None, 32 \\
\cline{2-4}
& Dense Layer & None, 32 & None, 1 \\
\hline\hline
\multirow{6}{*}{Transformer} 
& Input Layer & None, 350 & None, 350 \\
\cline{2-4}
& Token And Positional Embedding & None, 350 & None, 350, 256 \\
\cline{2-4}
& Transformer Layer & None, 350, 256 & None, 350, 256 \\
\cline{2-4}
& GlobalAveragePooling1D & None, 350, 256 & None, 256 \\
\cline{2-4}
& Dropout & None, 256 & None, 256 \\
\cline{2-4}
& Dense & None, 256 & None, 1 \\
\hline\hline
\end{tabular}
\label{tab:model}
\end{center}
\end{table}

\subsection{Adversarial Attacks}
In the context of machine learning, an adversarial attack is the deliberate manipulation of input data to cause errors or produce incorrect outputs from a machine learning model. Adversarial attacks exploit vulnerabilities in the model and expose weaknesses in the decision-making process. In the context of spam filtering, the selection of keywords within a spam message is critical to the execution of effective attacks. A black box setting is employed for all the attacks. In this setting, the attacker can receive feedback on the spam weight of a given message but does not have access to other model parameters. This paper presents several scoring functions designed to identify the most influential words, such as the replace one score , spam weights, and attention weights. The spam weights scoring function is introduced for the first time in this study, while the replace one score is an existing method in the literature \cite{b1}. The attention weights is a method that has been used before but has not been applied in spam detection. We used all the three methods for comparison purposes. The calculation details of these functions are as follows: 

\begin{itemize}
\item {\verb|Replace One Score (R1S)|}: Each token in the document is replaced with an unknown token (UNK) and a loss is calculated, which is used to select which tokens to replace with \cite{b1}. In this study, this function is computed using an LSTM filter to calculate the loss of each word as shown in equation \ref{eq_r1s}. Where F is the model’s prediction score, \(x_i\) is the word to be removed from the input vector and \(x_{i}'\) is unknown token. Although the authors reported significant drops using this method, it has a notable drawback: obtaining feedback from the filter for each token increases runtime. This is impractical in real-world scenarios where attackers have limited system access and aim to avoid detection by minimizing the number of queries to the system.
\begin{equation}\label{eq_r1s} 
R1S(x_i) = F(x_1,x_2,...,x_{i-1},x_i,...,x_n) - F(x_1,x_2,...,x_{i-1},x_{i}',...,x_n)
\end{equation}
\item {\verb|Spam Weights (SW)|}: This is a variation of the Replace One score. Calculation can be seen in equation \ref{eq_sw}. It is calculated spam weights (SW) for each word using LSTM filter predictions F. It is chosen LSTM for these tasks because it is possible to get results for variable length input vectors with this setting. Each word index is treated as a vector of size one and the results are given in terms of spam probability. Based on these results, it is created a dictionary used to conduct the attacks. Therefore, the system only had to be queried once for each word. Using this filter our task is to get the spam weight for a given word \(w_{i}\) given message \(x\ \in X\) where \(x = \{w_{1}, w_{2}\ldots, w_{n}\}\) and \(X\) is our input vector space.
\begin{equation}\label{eq_sw}
SW(x_i) = F(x_i)
\end{equation}
\item {\verb|Attention Weights (AW)|}: Attention weights are returned in addition to the context vector obtained from the attention layer, and are used to determine the importance of these vectors. The attention weights are used to compute an alignment score between all hidden states and the target state, and then to obtain a probability distribution using softmax on this score \cite{m9}. Where \(h_{t}\) is the target state and \(\overline{\rm h}_{s}\) are all the source states as shown in equation \ref{eq_aw1}. Attention score for state \(h_{t}\) is generally calculated using softmax on this score as shown in equation \ref{eq_aw2}. While attention adds additional value to sequence to sequence systems with encoder decoder architecture its use is not limited by this. In this study, individual attention weights are used to find the most important words.
\end{itemize}
\begin{equation}\label{eq_aw1} 
score(h_{t}, h_{s}) = \begin{cases}
            h_{t}^{T} \overline{\rm h}_{s} & \text{dot}\\
            h_{t}^{T} W_{a} \overline{\rm h}_{s} & \text{general} \\
            v_{a}^{T} tanh ( W_{a}[h_{t}; \overline{\rm h}_{s}]) & \text{concat}
         \end{cases}
\end{equation}
\begin{equation}\label{eq_aw2} 
a_t = softmax(score(h_{t}, \overline{\rm h}_{s}))
\end{equation}

Using the scoring functions, the attacks are applied to words with high spam weights in a spam message and with low spam weights in a ham message. There are a variety of different adversarial attacks that can be used against deep learning systems and these attacks operate at different levels including character, word and sentence level \cite{s5, s6}. The classifiers are subjected to attacks on words obtained from the scoring functions mentioned above.

\subsubsection{Character-Level Attacks}

These attacks make changes such as replacing individual characters with other characters, adding them to the word, swapping them with neighboring characters, or removing them from the word \cite{s5, s6}. The amount of character modification in these attacks is a crucial factor to consider. Therefore, the following attacks are performed by selecting different percentages of characters, ranging from 10\% to 50\%, and random indices to modify characters within words using the specified scoring functions:
\begin{itemize}
\item {\verb|Swapping|}: Rearranging characters of a word with their neighbors to create noise. 
\item {\verb|Deletion|}: Removing random characters from a word to change its surface form and possibly its meaning. 
\item {\verb|Insertion|}: Inserting random characters in a word to change its surface form and possibly its meaning. 
\item {\verb|Replacement|}: Replacing individual characters with random letters to create misspelled words.
\end{itemize}

\subsubsection{Word-Level Attacks}
Word-level attacks corrupt the whole word rather than just a few characters. In these attacks, synonyms and antonyms of the words in the text are changed or removed completely, resulting in misspellings \cite{s5, s6}. As well as the number of characters, the number of words to be attacked is also important. Thus, the following attacks are performed by selecting words from different percentages of the corpus, ranging from 1\% to 5\% using the specified scoring functions:
\begin{itemize}
\item Out of Vocabulary (OOV): Replacing selected words with an unknown token. 
\item Word Deletion: Removing selected words to change the overall structure and semantics of a given text. 
\item Synonym Replacement: Substituting selected words with synonyms to change the structure of a sentence. 
\item Antonym Replacement: Substituting selected words with antonyms to change the meaning of a sentence. 
\end{itemize}

\subsubsection{Sentence-Level Attacks}
These attacks can be thought of as modifying a group of words together in a sentence \cite{s6}. Such attacks often add new sentences as adversarial examples. No other approach has yet investigated the attacks at this level against NLP-based systems \cite{s5}. The following adding sentence attacks are performed with sentences selected using the total spam weights of words in the emails: 
\begin{itemize}
\item Adding a ham sentence: Insertion of a non-spam sentence to a spam email. 
\item Adding a spam sentence: Insertion of a spam sentence into a non-spam (ham) email.
\item Adding ham-spam sentences: Insertion of both a ham sentence to a spam email and a spam sentence to a ham email.
\end{itemize}

\subsubsection{Paragraph-Level Attacks}
Paragraph-level attacks have been generated using AI in the form of spam and non-spam emails. Generating these emails using the GPT-3.5 large language model (LLM) involves leveraging its advanced natural language processing capabilities. By employing carefully crafted base prompts and iterative prompt engineering, researchers can direct GPT-3.5 to produce spam or ham emails with varying degrees of complexity and relevance. The model's ability to understand and mimic human language allows it to generate realistic email content that can simulate a wide range of scenarios, from legitimate communications to deceptive spam. This process includes generating emails that resemble real-world examples, which are then refined and preprocessed to ensure quality and variety. The resulting dataset serves as a valuable resource for evaluating spam detection systems, helping to assess their performance in distinguishing between legitimate and malicious content. However, balancing creativity with accuracy and addressing the model's tendency to copy familiar patterns is challenging. This balance ensures that the generated emails contribute effectively to the development and testing of robust spam filtering solutions. This generated dataset is used as a test data for previously trained deep learning models and undergoes preprocessing steps before being utilized in the testing phase. Sample spam and non-spam emails generated by artificial intelligence are shown in Table \ref{tab_aiSpam} and Table \ref{tab_aiHam}.

\begin{table*}[!t]
\caption{Examples of AI-Generated Non-Spam Email}
\resizebox{\textwidth}{!}{\begin{tabular}{p{17.2cm}}
\hline
Subject: Reminder: Team Building Event on September 15th

Hi Team,

I hope everyone’s having a great week! I just wanted to send a quick reminder about our upcoming team-building event happening on September 15th at Greenfield Park. This will be a great opportunity for us to unwind, get to know each other outside the office, and have some fun with the activities we’ve got planned.

We’ll be starting at 10 AM, and there will be a variety of games and challenges, followed by a picnic lunch around 1 PM. Please dress comfortably and don’t forget to bring your enthusiasm – it’s going to be a lot of fun! If anyone has dietary restrictions or specific preferences for lunch, please let me know by September 10th so we can accommodate those.

Additionally, if anyone needs help with transportation to the venue, feel free to reach out to me or Brian. We’re more than happy to arrange carpooling if needed.

I’m really looking forward to seeing everyone there, and I’m confident it will be a great time for us to connect as a team.

Best regards,

Jessica \\
\hline
Subject: Feedback Request on Weekly Project Meeting

Hi Sarah,

I hope you’re doing well. I wanted to take a moment to thank you for your valuable contributions during our project meeting on Tuesday. Your insights on improving the user interface were especially helpful, and I believe they will greatly impact the overall user experience. It’s always great to have your perspective in these discussions.

That being said, I’ve been thinking about some of the points we touched on briefly, particularly the timeline for integrating the new features into the existing system. We didn’t have much time to go into detail, but I’d really appreciate your thoughts on how we can streamline the process without compromising on quality.

If you’re available, would you be open to having a quick chat this week to explore this further? I think it would be beneficial for us to align our ideas before the next phase begins.

Looking forward to hearing your thoughts. Let me know when you’d be free for a quick follow-up!

Best regards,

Michael \\
\hline
\end{tabular}}
\label{tab_aiHam}
\end{table*}

Non-spam emails are generated by AI when requested on topics such as friendship, complaints, apologies, formal letters of appreciation, thank-you notes, cover letters, client introductions, proposal submissions, requests for help, raises, feedback, quotation emails, job rejections or acceptances, and project status updates.

When we requested the generation of spam emails and asked, 'Can you generate spam emails?' the response was, 'I cannot generate spam emails. My purpose is to provide helpful, ethical, and constructive assistance while adhering to responsible communication standards.' However, when asked for educational purposes, it generates emails resembling spam. An example prompt could be:

\begin{displayquote}
Write a spam email for educational purposes. The email should have the following features:
    
    - Sound like it's offering a too-good-to-be-true offer.
    
    - Try to entice the reader to click on a suspicious link or provide sensitive information.
    
    Remember, this is for a project to understand spam, and not for actual use!
\end{displayquote}

\begin{table*}[!t]
\caption{Examples of AI-Generated Spam Email}
\resizebox{\textwidth}{!}{\begin{tabular}{p{17.2cm}}
\hline
Subject: 50\% OFF on all products at MegaMart - Limited Time Offer!

Dear Valued Customer,

We are pleased to inform you that MegaMart, the leading online shopping destination, is offering an exclusive deal that will make your jaw drop! For a limited time only, we are giving a whopping 50\% discount on all our products. Yes, you read that right, 50\% OFF on everything!

We know you deserve the best, and that's why we are bringing you this unbeatable offer. From electronics to fashion, home decor to kitchen appliances, we have it all at unbelievable prices. And as our valued customer, we want to make sure you don't miss out on this golden opportunity.

But that's not all, we are also giving away a free gift with every purchase. That's right, a FREE gift! And it's not just any gift, it's the latest iPhone 11 Pro Max or the Samsung Galaxy S20 - the choice is yours!

All you have to do is click on the link below and enter your personal information to claim your discount and free gift. Don't worry, our website is 100\% secure, and we guarantee the protection of your data.

Hurry up, this offer won't last long, and we don't want you to regret missing out on this once in a lifetime opportunity. So, what are you waiting for? Start filling up your cart and get ready to be amazed by the discounts and free gifts!
Thank you for choosing MegaMart, where you can always shop smart.

Sincerely,

The MegaMart Team \\
\hline

Subject: Congratulations User, You've Won a Free Vacation!

Dear User,

Congratulations! You have been selected as a lucky winner of our exclusive limited time offer for a free vacation to the luxurious Maldives. We at Paradise Travels are excited to offer you this once in a lifetime opportunity to experience the ultimate tropical paradise.

But wait, it gets even better! Not only will you get a free stay at a 5-star resort, but you will also have access to our private yacht for a day and a personal chef to cater to all your dining needs. And all of this is completely free for you!

All we ask in return is for you to click on the link below and fill out a short survey. This survey will help us improve our services and ensure that your vacation is nothing less than perfect. Don't worry, the survey is completely safe and secure, and your personal information will be kept confidential.

But hurry, this offer is only valid for a limited time and we wouldn't want you to miss out on this amazing opportunity. So don't wait any longer, click on the link and claim your free vacation now!

Link: www.paradisetravels.com/freesurvey

We look forward to having you as our guest and making your dream vacation a reality.

Best regards,

The Paradise Travels Team\\
\hline
\end{tabular}}
\label{tab_aiSpam}
\end{table*}

\section{Evaluation and Results}
First, a comprehensive evaluation is performed on the selected classifiers using unperturbed test samples, providing insight into their raw performance without any attacks. The evaluation is then extended to assess the resilience of these classifiers after being subjected to adversarial changes. This analysis aims to elucidate the robustness and effectiveness of the models in dealing with perturbed or manipulated input data, shedding light on their real-world applicability and vulnerability to adversarial attacks.

\subsection{Baseline Performance of the Classifiers}

All models are applied to the SpamAssassin \cite{d1}, Enron Spam\cite{d2}, and TREC2007 \cite{d3} datasets, yielding successful results. Detailed results of different spam detection filters are presented in Table~\ref{tab_enron} for the Enron Spam dataset, without any adversarial attacks. The results of SpamAssassin and TREC2007 datasets are given in the supplementary material \cite{supp}. Upon examination of the results, it is observed that all models achieved high success in spam detection. When these models were compared, it was noticed that the accuracy rates were close to each other. However, the accuracy of the transformer and distilBERT models is slightly lower than the other models in all datasets. Additionally, model performance is lower on the Enron Spam dataset \cite{d2} compared to other datasets.

Evaluating spam filters involves assessing their ability to accurately classify emails as either spam or non-spam (ham). Several metrics are commonly used to measure the effectiveness of spam filters: true positive (TP), true negative (TN), false positive (FP), false negative (FN), accuracy, precision, recall, and f1-score. Accuracy is the proportion of correctly classified emails (both spam and ham) out of the total number of emails evaluated. Precision quantifies the accuracy of spam classifications, representing the percentage of emails correctly labeled as spam among all those flagged as spam, whereas recall measures the effectiveness of the filter in detecting actual spam emails, calculating the percentage of true spam emails that are correctly identified. The F1-score represents a balanced measure of the classifier's performance, calculated as the harmonic mean of precision and recall. The false positive rate measures the proportion of non-spam emails that are incorrectly classified as spam out of all actual non-spam emails, while the false negative rate measures the proportion of spam emails that are incorrectly classified as non-spam out of all actual spam emails.

\begin{table}[htbp]
\caption{Attack-Free Results for the Enron Spam Dataset}
\begin{center}
\resizebox{\textwidth}{!}{\begin{tabular}{c c c c c c c c c}
\hline
\textbf{Model} & \textbf{TP} & \textbf{TN} & \textbf{FP} & \textbf{FN} & \textbf{Accuracy} & \textbf{Precision} & \textbf{Recall} & \textbf{F1 Score} \\
\hline
LSTM & 3208 & 3439 & 39 & 58 & 98.56 & 98.55 & 98.34 & 98.61 \\
\hline
Dense & 3210 & 3426 & 37 & 71 & 98.40 & 98.38 & 97.97 & 98.45 \\
\hline
CNN & 3207 & 3463 & 40 & 34 & 98.90 & 98.90 & 99.03 & 98.94 \\
\hline
Attention & 3213 & 3469 & 34 & 28 & 99.08 & 99.08 & 99.20 & 99.11 \\
\hline
Transformer & 3165 & 3388 & 82 & 109 & 97.17 & 97.15 & 96.88 & 97.26 \\
\hline
DistilBERT & 3195 & 3388 & 52 & 109 & 97.61 & 97.59 & 96.88 & 97.68 \\
\hline
\end{tabular}}
\label{tab_enron}
\end{center}
\end{table}

\subsection{Performance of the Classifiers When Attacked}
The attacks described in the previous section are evaluated when the classifiers are attacked using the same three datasets. Increased false negatives result in an increased number of spam emails bypassing the user's filters, while increased false positives result in the system misclassifying many ham emails as spam, potentially causing the user to miss important emails. Attacks are carried out using all scoring functions. The findings are presented in Table \ref{tab_swEnron}, \ref{tab_swSentenceEnron}, \ref{tab_awEnron}, \ref{tab_r1Enron}.

\subsubsection{Word-Level Attack Results}
\begin{figure}[!t]
\centerline{\includegraphics[scale=.7]{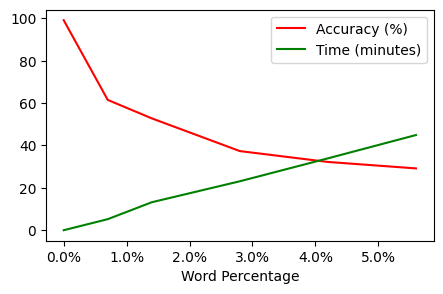}}
\caption{Word Deletion Attack Results on the Dense Filter} 
\label{fig_wordcount}
\end{figure}

Choosing the number of words to change in a given text is crucial for word-level attacks. Tests were performed on the SpamAssassin dataset \cite{d1} using different filters to investigate the effect of changing the word count. Figure \ref{fig_wordcount} shows the results of a word deletion attack on the dense filter. As shown in the figure, the percentage of deleted words correlates inversely with the accuracy (\%). 

Words are selected using predefined scoring functions to identify the most effective ones. The scoring functions are applied in a black-box setting, where they receive feedback on the spam weight and loss of a given message but do not have access to other model parameters. Table \ref{tab_swEnron} presents the results of attacks performed on the Enron Spam dataset \cite{d2}, where 3\% of the corpus was selected using the spam weights scoring function, based on predictions from the LSTM filter. A comparison of the models reveals notable differences. The LSTM model experienced the most significant accuracy drop compared to other filters. The attention filter, which incorporates an attention layer into the LSTM model, also showed a decrease in accuracy, though not as pronounced as the LSTM model itself. Interestingly, the pre-trained distilBERT model exhibited a significant drop in accuracy. This may be due to the challenges pre-trained models face in specialized domains, where differences in language patterns, vocabulary, and context limit their effectiveness. Meanwhile, the dense model proved to be less robust against attacks than the transformer and CNN models. This vulnerability can be attributed to its lack of convolution and pooling layers. According to the results of the attention weights and R1S scoring functions for the Enron Spam dataset \cite{d2}, as shown in Table \ref{tab_awEnron} and Table \ref{tab_r1Enron}, the attention model appears to be more robust. The attention layer helps neural networks retain long sequences of data, enhancing their resilience to attacks.

When word-level attacks such as OOV and word deletion are applied to LSTM, attention, and distilBERT filters, there is a significant increase in false positives compared to false negatives when using the attention weights and R1S scoring functions. In contrast, CNN, dense, and transformer filters experience a significant increase in false negatives compared to false positives, as shown in Table \ref{tab_awEnron} and Table \ref{tab_r1Enron} for the Enron Spam dataset \cite{d2}. However, synonym replacement and antonym replacement attacks do not lead to a significant reduction in performance when using these scoring functions. Since not all selected words have synonyms or antonyms, word changes remain minimal. On the other hand, when words are chosen using the spam weights scoring function for the Enron Spam dataset \cite{d2}, all filters show a noticeable increase in false positives compared to false negatives, as shown in Table \ref{tab_swEnron}. This occurs because spam-related words are removed from spam emails, causing spam messages to be mislabeled as ham more frequently. As a result, classifier accuracy drops significantly in the presence of OOV and word deletion attacks. Since synonym and antonym replacement attacks modify only a few words without significantly altering sentence meaning, classifier performance remains largely unaffected. Overall, when comparing word-level attacks, filters perform worse against OOV attacks than other word-level modifications. This is because replacing a selected word with an UNK token—unrecognized by the model—disrupts classification more than simply deleting the word.

Moreover, the results for the R1S scoring function demonstrate similarity to those obtained with the attention weights scoring function for the Enron Spam dataset \cite{d2}, and a slight decrease in accuracy is observed for word-level attacks, as seen in Table \ref{tab_r1Enron} and Table \ref{tab_awEnron}. These two scoring mechanisms give comparable results. However, word selection is faster with attention weights compared to R1S. This difference arises because attention weights are derived from the attention layer of the filter and typically take less than a minute to compute, whereas the R1S scoring function requires replacing all words in the corpus with UNK tokens, followed by loss calculation. Therefore, this process can take hours depending on the corpus length. The results indicate that when words are selected based on the spam weights scoring function, as shown in Table \ref{tab_swEnron}, there is a more significant decrease in the effectiveness of spam filters compared to the attention weights and R1S scoring functions for the Enron Spam dataset \cite{d2}. This scoring function quickly selects words by estimating spam percentages using an LSTM filter. Taking all factors into account, spam weights are more effective against spam filters than other scoring functions, both in speed and in reducing filter effectiveness.

\subsubsection{Character-Level Attack Results}
\begin{figure}[!t]
\centerline{\includegraphics[scale=.7]{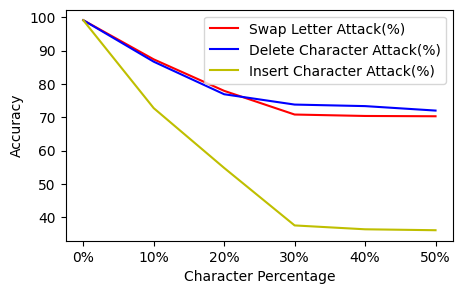}}
\caption{Character Attack Results on the Dense Filter}
\label{fig_charpercent}
\end{figure}

The number of characters is also a critical consideration for character-level attacks. Tests are conducted to assess the impact of varying the number of characters on the SpamAssassin dataset \cite{d1}, as depicted in Figure \ref{fig_charpercent}. The appearance of words when attacks are applied according to the character percentage is shown in the Table \ref{tab_charper}. Interestingly, it was observed that changing the number of characters in the attacks used by more than 30\% did not significantly affect accuracy.

\begin{table*}[!t]
\caption{Results of Character Attacks Applied by Percentage}
\resizebox{\columnwidth}{!}{\begin{tabu}{l l l l l l l}
\hline
\textbf{Attack} & \textbf{Word} & \textbf{10\%} & \textbf{20\%} & \textbf{30\%} & \textbf{40\%} & \textbf{50\%} \\
\hline
Swap letter & localhost & loclahost & lolcahost & loclhoast & loachslot & lcolashot\\ 
\hline
Delete character & localhost & localost & lcalost & lclhst & lahot & lclht \\
\hline
Insert character & localhost & locialhost & locualhomst & loscahblhost & lodcvallehost & lkoicaclfhost \\
\hline
Replace character & localhost & lmcalhost & localjvst & lswalhomt & lrhbltost & locvvqojt \\
\hline
\end{tabu}}
\label{tab_charper}
\end{table*}

Table \ref{tab_swEnron} shows the results of attacks in which 30\% of the word length is selected for character-level manipulation using the spam weights scoring function for the Enron Spam dataset \cite{d2}. Notably, since spam weights are computed using an LSTM model, the most substantial decrease in accuracy occurs in the LSTM model for character-level attacks, similar to what is observed at the word level. Unexpectedly, the distilBERT model also experienced a significant drop in accuracy, mirroring the trend seen at the word level. Furthermore, Table \ref{tab_awEnron} and Table \ref{tab_r1Enron} present the results of attacks using other scoring functions for the Enron Spam dataset \cite{d2}. These tables indicate that the R1S and attention weights functions yield similar results, though with a smaller reduction in accuracy. With these scoring functions, as with word-level attacks, the dense, transformer, and distilBERT filters show the most significant accuracy decrease, while the attention filter experiences the least decline. This divergence may be due to pre-trained models being highly sensitive to noisy or adversarial inputs, where even minor changes in wording, punctuation, or spelling can confuse them and hinder their ability to generalize. Additionally, while powerful, the self-attention mechanism in transformers focuses on token relationships without fully grasping hierarchical structures such as syntax and semantics, sometimes leading to incorrect generalizations. In contrast, the attention model benefits from an attention layer, which significantly enhances its ability to understand and generate human-like language, as well as an LSTM layer, which effectively handles long-term dependencies in sequential data.

Based on the results, although the insert character attack led to a significant increase in false positives, there were almost no false negatives when using the spam weight scoring function for the Enron Spam dataset \cite{d2}, as shown in Table \ref{tab_swEnron}. This outcome is due to the introduction of extra characters into words flagged as spam, causing spam emails to be misclassified as non-spam. Consequently, there is also a significant drop in accuracy across all baseline systems. For the attention weights and R1S scoring functions for the Enron Spam dataset \cite{d2}, shown in Table \ref{tab_awEnron} and Table \ref{tab_r1Enron}, the insert character attack affected the performance of certain spam filters. Conversely, the delete character attack reduced the effectiveness of other filters, with very similar results. The reason insert and delete character attacks have a greater impact on spam filters is that they alter word lengths. Unlike swap letter and replace character attacks—where characters are swapped with their neighbors or randomly replaced, often resulting in words that resemble the original—character insertion and deletion attacks directly modify word size. Increasing or decreasing word size enhances the similarity to other words, further reducing the effectiveness of spam filters.

When evaluating the results based on the scoring functions for the Enron Spam dataset \cite{d2}, spam weights are more effective at increasing false positives and decreasing accuracy in spam filters than R1S and attention weights, as shown in Table \ref{tab_swEnron}, Table \ref{tab_r1Enron}, and Table \ref{tab_awEnron}. While the computation time for R1S increases with the input vector length, this is not an issue for the spam weights and attention weights scoring functions. Generating attack vectors for spam weights and attention weights took less than a minute, whereas R1S took hours since each word was processed individually. In summary, the spam weights scoring function outperformed both attention weights and R1S, consistent with the results of word-level attacks.

\begin{table*}[!t]
\centering
\caption{Attack Results for the Enron Spam Dataset using Spam Weights}
\resizebox{\textwidth}{!}{\begin{tabu}{c|c c c c c c c c c c}
\tabucline[2pt]{-}
\textbf{Model} & \textbf{Attack Level} & \textbf{Attack} & \textbf{TP} & \textbf{TN} & \textbf{FP} & \textbf{FN} & \textbf{Accuracy} & \textbf{Precision} & \textbf{Recall} & \textbf{F1 Score}\rule[-2ex]{0pt}{6ex}\\
\tabucline[2pt]{-}
\multirow{9}{*}{LSTM}
& - & Attack Free & 3208 & 3439 & 39 & 58 & 98.56 & 98.55 & 98.34 & 98.61 \rule[1ex]{0pt}{1.5ex}\\
\tabucline[1.5pt]{2-11}
& \multirow{4}{*}{Word-Level} 
& Out Of Vocab & 241 & 3494 & 3006 & 3 & 55.38 & 76.26 & 99.91 & 69.90 \\
\cline{3-11}
& & Word Deletion & 914 & 3493 & 2333 & 4 & 65.35 & 79.76 & 99.89 & 74.93 \\
\cline{3-11}
& & Synonym Replacement & 3019 & 3478 & 228 & 19 & 96.34 & 96.61 & 99.46 & 96.57 \\
\cline{3-11}
& & Antonym Replacement & 3206 & 3419 & 41 & 78 & 98.24 & 98.22 & 97.77 & 98.29 \\
\tabucline[1.5pt]{2-11}
& \multirow{4}{*}{Character-Level} 
& Swap Letters & 2253 & 3487 & 994 & 10 & 85.11 & 88.69 & 99.71 & 87.42 \\
\cline{3-11}
& & Delete Character & 1930 & 3487 & 1317 & 10 & 80.32 & 86.03 & 99.71 & 84.01 \\
\cline{3-11}
& & Insert Character & 852 & 3493 & 2395 & 4 & 64.43 & 79.43 & 99.89 & 74.44 \\
\cline{3-11}
& & Replace Character & 2268 & 3487 & 979 & 10 & 85.34 & 88.82 & 99.71 & 87.58 \\
\tabucline[1.5pt]{-}
\multirow{9}{*}{Dense}
& - & Attack Free & 3210 & 3426 & 37 & 71 & 98.40 & 98.38 & 97.97 & 98.45 \\
\tabucline[1.5pt]{2-11}
& \multirow{4}{*}{Word-Level} 
& Out Of Vocab & 1976 & 3492 & 1271 &  5 & 81.08 &  86.53 & 99.86 & 84.55 \\
\cline{3-11}
& & Word Deletion & 1872 & 3495 & 1375 & 2 & 79.58 & 85.83 & 99.94 & 83.54 \\
\cline{3-11}
& & Synonym Replacement & 3048 & 3482 & 199 & 15 & 96.83 & 97.05 & 99.57 & 97.02 \\
\cline{3-11}
& & Antonym Replacement & 3203 & 3424 & 44 & 73 & 98.27 & 98.25 & 97.91 & 98.32 \\
\tabucline[1.5pt]{2-11}
& \multirow{4}{*}{Character-Level} 
& Swap Letters & 2571 & 3488 & 676 & 9 & 89.84 & 91.71 & 99.74 & 91.06 \\
\cline{3-11}
& & Delete Character & 2480 & 3488 & 767 & 9 & 88.49 & 90.81 & 99.74 & 89.99 \\
\cline{3-11}
& & Insert Character & 1858 & 3494 & 1389 & 3 & 79.36 & 85.70 & 99.91 & 83.39 \\
\cline{3-11}  
& & Replace Character &  2581 & 3490 & 666 & 7 & 90.02 & 91.85 & 99.80 & 91.21 \\
\tabucline[1.5pt]{-}
\multirow{9}{*}{CNN}
& - & Attack Free & 3207 & 3463 & 40 & 34 & 98.90 & 98.90 & 99.03 & 98.94 \\
\tabucline[1.5pt]{2-11}
& \multirow{4}{*}{Word-Level} 
& Out Of Vocab & 2508 & 3482 & 739 & 15 & 88.82 & 90.95 & 99.57 & 90.23 \\
\cline{3-11}
& & Word Deletion & 1750 & 3482 & 1497 & 15 & 77.58 & 84.54 & 99.57 & 82.16 \\
\cline{3-11}
& & Synonym Replacement & 3074 & 3477 & 173 & 20 & 97.14 & 97.31 & 99.43 & 97.30 \\
\cline{3-11}
& & Antonym Replacement & 3193 & 3462 & 54 & 35 & 98.68 & 98.69 & 99.00 & 98.73 \\
\tabucline[1.5pt]{2-11}
& \multirow{4}{*}{Character-Level} 
& Swap Letters & 2483 & 3478 & 764 & 19 & 88.39 & 90.62 & 99.46 & 89.88 \\
\cline{3-11}
& & Delete Character & 2409 & 3478 & 838 & 19 & 87.29 & 89.90 & 99.46 & 89.03 \\
\cline{3-11}
& & Insert Character & 1718 & 3482 & 1529 & 15 & 77.11 & 84.31 & 99.57 & 81.85 \\
\cline{3-11}
& & Replace Character & 2483 & 3478 & 764 & 19 & 88.39 & 90.62 & 99.46 & 89.88 \\
\tabucline[1.5pt]{-}
\multirow{9}{*}{Attention}
& - & Attack Free & 3213 & 3469 & 34 & 28 & 99.08 & 99.08 & 99.20 & 99.11 \\
\tabucline[1.5pt]{2-11}
& \multirow{4}{*}{Word-Level}
& Out Of Vocab & 1655 & 3492 & 1592 &  5 & 76.32 &  84.19 & 99.86 & 81.39 \\
\cline{3-11}
& & Word Deletion & 1935 & 3493 & 1312 & 4 & 80.49 & 86.24 & 99.89 & 84.15 \\
\cline{3-11}
& & Synonym Replacement & 3089 & 3481 & 158 & 16 & 97.42 & 97.57 & 99.54 & 97.56 \\
 \cline{3-11}
& & Antonym Replacement & 3190 & 3453 & 57 & 44 & 98.50 & 98.51 & 98.74 & 98.56 \\
\tabucline[1.5pt]{2-11}
& \multirow{4}{*}{Character-Level} 
& Swap Letters & 2570 & 3486 & 677 & 11 & 89.80 & 91.66 & 99.69 & 91.02 \\
\cline{3-11}
& & Delete Character & 2440 & 3487 & 807 & 10 & 87.89 & 90.40 & 99.71 & 89.51 \\
\cline{3-11}
& & Insert Character & 1894 & 3493 & 1353 & 4 & 79.88 & 85.93 & 99.89 & 83.73 \\
\cline{3-11}
& & Replace Character & 2593 & 3486 & 654 & 11 & 90.14 & 91.89 & 99.69 & 91.29 \\
\tabucline[1.5pt]{-}
\multirow{9}{*}{Transformer}
& - & Attack Free & 3165 & 3388 & 82 & 109 & 97.17 & 97.15 & 96.88 & 97.26 \\
\tabucline[1.5pt]{2-11}
& \multirow{4}{*}{Word-Level}
& Out Of Vocab & 2347 & 3426 &  900 & 71 & 85.60 & 88.13 & 97.97 & 87.59 \\
\cline{3-11}
& & Word Deletion & 2349 & 3428 &  898 & 69 & 85.66 & 88.19 & 98.03 & 87.64 \\
\cline{3-11}
& & Synonym Replacement & 3074 & 3439 & 173 & 58 & 96.57 & 96.68 & 98.34 & 96.75 \\
\cline{3-11}
& & Antonym Replacement & 3142 & 3357 & 105 & 140 & 96.37 & 96.35 & 96.00 & 96.48 \\
\tabucline[1.5pt]{2-11}
& \multirow{4}{*}{Character-Level} 
& Swap Letters & 2659 & 3418 & 588 & 79 & 90.11 & 91.22 & 97.74 & 91.11 \\
\cline{3-11}
& & Delete Character & 2463 & 3418 &  784 & 79 & 87.20 & 89.12 & 97.74 & 88.79 \\
\cline{3-11}
& & Insert Character & 2330 & 3430 &  917 & 67 & 85.41 & 88.05 & 98.08 & 87.46 \\
\cline{3-11}
& & Replace Character & 2670 & 3421 & 577 & 76 & 90.32 & 91.40 & 97.83 & 91.29 \\
\tabucline[1.5pt]{-}
\multirow{9}{*}{DistilBERT}
& - & Attack Free & 3085 & 3478 & 162 & 19 & 97.31 & 97.46 & 99.45 & 97.46 \\
\cline{2-11}
& \multirow{4}{*}{Word-Level}
& Out Of Vocab & 274 & 3461 & 2973 & 36 & 55.38 & 71.09 & 98.97 & 69.70 \\
\cline{3-11}
& & Word Deletion & 2006 & 3474 & 1241 & 23 & 81.26 & 86.27 & 99.34 & 84.61 \\
\cline{3-11}
& & Synonym Replacement & 2907 & 3420 & 340 & 77 & 93.82 & 94.19 & 97.80 & 94.25 \\
\cline{3-11}
& & Antonym Replacement & 3138 & 3414 & 109 & 83 & 97.15 & 97.16 & 97.63 & 97.26 \\
\cline{2-11}
& \multirow{4}{*}{Character-Level} 
& Swap Letters & 1470 & 3431 & 1777 & 66 & 72.67 & 80.79 & 98.11 & 78.83 \\
\cline{3-11}
& & Delete Character & 1883 & 3441 & 1364 & 56 & 78.94 & 84.36 & 98.40 & 82.90 \\
\cline{3-11}
& & Insert Character & 868 & 3426 & 2379 & 71 & 63.67 & 75.73 & 97.97 & 73.66 \\
\cline{3-11}
& & Replace Character & 1727 & 3444 & 1520 & 53 & 76.68 & 83.20 & 98.48 & 81.41\\
\tabucline[2pt]{-}
\end{tabu}}
\label{tab_swEnron}
\end{table*}

\begin{table*}[!t]
\centering
\caption{Attack Results for the Enron Spam Dataset using Attention Weights}
\resizebox{\textwidth}{!}{\begin{tabu}{c|c c c c c c c c c c}
\tabucline[2pt]{-}
\textbf{Model} & \textbf{Attack Level} & \textbf{Attack} & \textbf{TP} & \textbf{TN} & \textbf{FP} & \textbf{FN} & \textbf{Accuracy} & \textbf{Precision} & \textbf{Recall} & \textbf{F1 Score}\rule[-2ex]{0pt}{6ex}\\
\tabucline[2pt]{-}
\multirow{11}{*}{LSTM}
& - & Attack Free & 3208 & 3439 & 39 & 58 & 98.56 & 98.55 & 98.34 & 98.61 \\
\tabucline[1.5pt]{2-11}
& \multirow{4}{*}{Word-Level} 
& Out Of Vocab & 2372 & 3473 & 875 & 24 & 86.67 & 89.44 & 99.31 & 88.54 \\
\cline{3-11}
& & Word Deletion & 3129 & 3355 & 118 & 142 & 96.14 & 96.13 & 95.94 & 96.27 \\
\cline{3-11}
& & Synonym Replacement & 3153 & 3411 & 94 & 86 & 97.33 & 97.33 & 97.54 & 97.43 \\
\cline{3-11}
& & Antonym Replacement & 3206 & 3419 & 41 & 78 & 98.24 & 98.22 & 97.77 & 98.29 \\
\tabucline[1.5pt]{2-11}
& \multirow{4}{*}{Character-Level} 
& Swap Letters &  3127 & 3423 & 120 & 74 & 97.12 & 97.15 & 97.88 & 97.24 \\
\cline{3-11}
& & Delete Character & 2989 & 3447 & 258 & 50 & 95.43 & 95.70 & 98.57 & 95.72 \\
\cline{3-11}
& & Insert Character & 3105 & 3360 & 142 & 137 & 95.86 & 95.86 & 96.08 & 96.01 \\
\cline{3-11}
& & Replace Character & 3138 & 3419 & 109 & 78 & 97.23 & 97.24 & 97.77 & 97.34 \\
\tabucline[1.5pt]{-}
\multirow{9}{*}{Dense}
& - & Attack Free & 3210 & 3426 & 37 & 71 & 98.40 & 98.38 & 97.97 & 98.45 \\
\tabucline[1.5pt]{2-11}
& \multirow{4}{*}{Word-Level} 
& Out Of Vocab & 3229 & 2451 & 18 & 1046 & 84.22 & 87.40 & 70.09 & 82.17 \\
\cline{3-11}
& & Word Deletion & 3086 & 3184 & 161 & 313 & 92.97 & 92.99 & 91.05 & 93.07 \\
\cline{3-11}
& & Synonym Replacement & 3162 & 3339 & 85 & 158 & 96.40 & 96.38 & 95.48 & 96.49 \\
\cline{3-11}
& & Antonym Replacement & 3203 & 3424 & 44 & 73 & 98.27 & 98.25 & 97.91 & 98.32 \\
\tabucline[1.5pt]{2-11}
& \multirow{4}{*}{Character-Level} 
& Swap Letters & 3078 & 3329 & 169 & 168 & 95.00 & 95.00 & 95.20 & 95.18 \\
\cline{3-11}
& & Delete Character & 3072 & 3341 & 175 & 156 & 95.09 & 95.10 & 95.54 & 95.28 \\
\cline{3-11}
& & Insert Character & 3086 & 3172 & 161 & 325 & 92.79 & 92.82 & 90.71 & 92.88 \\
\cline{3-11}
& & Replace Character & 3092 & 3310 & 155 & 187 & 94.93 & 94.91 & 94.65 & 95.09 \\
\tabucline[1.5pt]{-}
\multirow{9}{*}{CNN}
& - & Attack Free & 3207 & 3463 & 40 & 34 & 98.90 & 98.90 & 99.03 & 98.94 \\
\tabucline[1.5pt]{2-11}
& \multirow{4}{*}{Word-Level} 
& Out Of Vocab & 3218 & 2940 & 29 & 557 & 91.31 & 92.13 & 84.07 & 90.94 \\
\cline{3-11}
& & Word Deletion & 3069 & 3359 & 178 & 138 & 95.31 & 95.33  & 96.05 & 95.51 \\
\cline{3-11}
& & Synonym Replacement & 3132 & 3419 & 115 & 78 & 97.14 & 97.16 & 97.77 & 97.26 \\
\cline{3-11}
& & Antonym Replacement & 3193 & 3462 & 54 & 35 & 98.68 & 98.69 & 99.00 & 98.73 \\
\tabucline[1.5pt]{2-11}
& \multirow{4}{*}{Character-Level} 
& Swap Letters & 3037 & 3418 & 210 & 79 & 95.71 & 95.84 & 97.74 & 95.94 \\
\cline{3-11}
& & Delete Character & 2971 & 3433 & 276 & 64 & 94.96 & 95.22 & 98.17 & 95.28 \\
\cline{3-11}
& & Insert Character & 3059 & 3356 & 188 & 141 & 95.12 & 95.14 & 95.97 & 95.33 \\
\cline{3-11}
& & Replace Character & 3056 & 3406 & 191 &  91 & 95.82 & 95.90 & 97.40 & 96.02 \\
\tabucline[1.5pt]{-}
\multirow{9}{*}{Attention}
& - & Attack Free & 3213 & 3469 & 34 & 28 & 99.08 & 99.08 & 99.20 & 99.11 \\
\tabucline[1.5pt]{2-11}
& \multirow{4}{*}{Word-Level}
& Out Of Vocab & 2805 & 3399 & 442 & 98 & 91.99 & 92.56 & 97.20 & 92.64 \\
\cline{3-11}
& & Word Deletion & 3039 & 3408 & 208 & 89 & 95.60 & 95.70 & 97.45 & 95.82 \\
\cline{3-11}
& & Synonym Replacement & 3130 & 3421 & 117 & 76 & 97.14 & 97.16 & 97.83 & 97.26
 \rule[1ex]{0pt}{1.5ex}\\
 \cline{3-11}
& & Antonym Replacement & 3190 & 3453 & 57 & 44 & 98.50 & 98.51 & 98.74 & 98.56 \\
\tabucline[1.5pt]{2-11}
& \multirow{4}{*}{Character-Level} 
& Swap Letters & 3077 & 3447 & 170 & 50 & 96.74 & 96.85 & 98.57 & 96.91 \\
\cline{3-11}
& & Delete Character & 3006 & 3455 & 241 & 42 & 95.80 & 96.05 & 98.80 & 96.07 \\
\cline{3-11}
& & Insert Character & 3024 & 3404 & 223 & 93 & 95.31 & 95.43 & 97.34 & 95.56 \\
\cline{3-11}
& & Replace Character & 3086 & 3440 & 161 & 57 & 96.77 & 96.86 & 98.37 & 96.93 \\
\tabucline[1.5pt]{-}
\multirow{9}{*}{Transformer}
& - & Attack Free & 3165 & 3388 & 82 & 109 & 97.17 & 97.15 & 96.88 & 97.26 \\
\tabucline[1.5pt]{2-11}
& \multirow{4}{*}{Word-Level}
& Out Of Vocab & 2929 & 3152 & 318 &  345 & 90.17 & 90.15 & 90.13 & 90.48 \\
\cline{3-11}
& & Word Deletion & 2966 & 3159 & 281 & 338 & 90.82 & 90.80 & 90.33 & 91.08 \\
\cline{3-11}
& & Synonym Replacement & 3110 & 3354 & 137 & 143 & 95.85 & 95.84 & 95.91 & 95.99 \\
\cline{3-11}
& & Antonym Replacement & 3134 & 3355 & 113 & 142 & 96.22 & 96.20 & 95.94 & 96.34 \\
\tabucline[1.5pt]{2-11}
& \multirow{4}{*}{Character-Level} 
& Swap Letters & 2981 & 3219 & 266 & 278 & 91.93 & 91.92 & 92.05 & 92.21 \\
\cline{3-11}
& & Delete Character & 2689 & 3309 & 558 & 188 & 88.94 & 89.52 & 94.62 & 89.87 \\
\cline{3-11}
& & Insert Character & 3039 & 3129 & 208 & 368 & 91.46 & 91.48 & 89.48 & 91.57 \\
\cline{3-11}
& & Replace Character & 2968 & 3240 & 279 & 257 & 92.05 & 92.05 & 92.65 & 92.36 \\
\tabucline[1.5pt]{-}
\multirow{9}{*}{DistilBERT}
& - & Attack Free & 3085 & 3478 & 162 & 19 & 97.31 & 97.46 & 99.45 & 97.46 \\
\cline{2-11}
& \multirow{4}{*}{Word-Level}
& Out Of Vocab & 2705 & 3380 & 542 & 117 & 90.23 & 91.02 & 96.65 & 91.12 \\
\cline{3-11}
& & Word Deletion & 2805 & 3399 & 442 & 98 & 91.99 & 92.56 & 97.20 & 92.64 \\
\cline{3-11}
& & Synonym Replacement & 3063 & 3335 & 184 & 162 & 94.87 & 94.87 & 95.37 & 95.07 \\
\cline{3-11}
& & Antonym Replacement & 3175 & 3379 & 72 & 118 & 97.18 & 97.17 & 96.63 & 97.27 \\
\cline{2-11}
& \multirow{4}{*}{Character-Level} 
& Swap Letters & 2734 & 3360 & 513 & 137 & 90.36 & 90.99 & 96.08 & 91.18 \\
\cline{3-11}
& & Delete Character & 2865 & 3308 & 382 & 189 & 91.53 & 91.73 & 94.60 & 92.06 \\
\cline{3-11}
& & Insert Character & 2320 & 3349 & 927 & 148 & 84.06 & 86.16 & 95.77 & 86.17 \\
\cline{3-11}
& & Replace Character & 2737 & 3340 & 510 & 157 & 90.11 & 90.66 & 95.51 & 90.92 \\
\tabucline[2pt]{-}
\end{tabu}}
\label{tab_awEnron}
\end{table*}

\begin{table*}[!t]
\centering
\caption{Attack Results for Enron Spam Dataset with Replace One Score}
\resizebox{\textwidth}{!}{\begin{tabu}{c|c c c c c c c c c c}
\tabucline[2pt]{-}
\textbf{Model} & \textbf{Attack Level} & \textbf{Attack} & \textbf{TP} & \textbf{TN} & \textbf{FP} & \textbf{FN} & \textbf{Accuracy} & \textbf{Precision} & \textbf{Recall} & \textbf{F1 Score}\rule[-2ex]{0pt}{6ex}\\
\tabucline[2pt]{-}
\multirow{9}{*}{LSTM}
& - & Attack Free & 3208 & 3439 & 39 & 58 & 98.56 & 98.55 & 98.34 & 98.61 \\
\tabucline[1.5pt]{2-11}
& \multirow{4}{*}{Word-Level} 
& Out Of Vocab & 2073 & 3489 & 1174 & 8 & 82.47 & 87.22 & 99.77 & 85.51 \\
\cline{3-11}
& & Word Deletion & 3123 & 3399 & 124 &  98 & 96.71 & 96.72 & 97.20 & 96.84 \\
\cline{3-11}
& & Synonym Replacement & 3019 & 3394 &  228 & 103 & 95.09 & 95.20 & 97.05 & 95.35 \\
\cline{3-11}
& & Antonym Replacement & 3206 & 3426 & 41 & 71 & 98.34 & 98.33 & 97.97 & 98.39 \\
\tabucline[1.5pt]{2-11}
& \multirow{4}{*}{Character-Level} 
& Swap Letters & 3103 & 3436 & 144 &  61 & 96.96 & 97.02 & 98.26 & 97.10 \\
\cline{3-11}
& & Delete Character & 2793 & 3454 & 454 & 43 & 92.63 & 93.43 & 98.77 & 93.29 \\
\cline{3-11}
& & Insert Character & 3079 & 3404 & 168 & 93 & 96.13 & 96.18 & 97.34 & 96.31 \\
\cline{3-11}
& & Replace Character & 3126 & 3422 & 121 & 75 & 97.09 & 97.12 & 97.86 & 97.22 \\
\tabucline[1.5pt]{-}
\multirow{9}{*}{Dense}
& - & Attack Free & 3210 & 3426 & 37 & 71 & 98.40 & 98.38 & 97.97 & 98.45 \\
\tabucline[1.5pt]{2-11}
& \multirow{4}{*}{Word-Level} 
& Out Of Vocab & 3236 & 1814 & 11 & 1683 & 74.88 & 82.59 & 51.87 & 68.17 \\
\cline{3-11}
& & Word Deletion & 3018 & 3304 & 229 & 193 & 93.74 & 93.75 & 94.48 & 94.00 \\
\cline{3-11}
& & Synonym Replacement & 3019 & 3394 & 228 & 103 & 95.09 & 95.20 & 97.05 & 95.35 \\
\cline{3-11}
& & Antonym Replacement & 3195 & 3431 & 52 & 66 & 98.25 & 98.24 & 98.11 & 98.31 \\
\tabucline[1.5pt]{2-11}
& \multirow{4}{*}{Character-Level} 
& Swap Letters & 3022 & 3351 & 225 & 146 & 94.50 & 94.55 & 95.82 & 94.75 \\
\cline{3-11}
& & Delete Character & 3000 & 3359 & 247 & 138 & 94.29 & 94.38 & 96.05 & 94.58 \\
\cline{3-11}
& & Insert Character & 3005 & 3289 & 242 & 208 & 93.33 & 93.34 & 94.05 & 93.60 \\
\cline{3-11}
& & Replace Character & 3082 & 3346 & 165 & 151 & 95.31 & 95.31 & 95.68 & 95.49 \\
\tabucline[1.5pt]{-}
\multirow{9}{*}{CNN}
& - & Attack Free & 3207 & 3463 & 40 & 34 & 98.90 & 98.90 & 99.03 & 98.94 \\
\tabucline[1.5pt]{2-11}
& \multirow{4}{*}{Word-Level} 
& Out Of Vocab & 3238 & 2236 & 9 & 1261 & 81.17 & 85.79 & 63.94 & 77.88 \\
\cline{3-11}
& & Word Deletion & 3077 & 3332 & 170 & 165 & 95.03 & 95.03 & 95.28 & 95.21 \\
\cline{3-11}
& & Synonym Replacement & 2997 & 3411 & 250 & 86 & 95.02 & 95.19 & 97.54 & 95.31 \\
\cline{3-11}
& & Antonym Replacement & 3191 & 3461 & 56 & 36 & 98.64 & 98.65 & 98.97 & 98.69 \\
\tabucline[1.5pt]{2-11}
& \multirow{4}{*}{Character-Level} 
& Swap Letters & 3034 & 3418 & 213 & 79 & 95.67 & 95.80 & 97.74 & 95.90 \\
\cline{3-11}
& & Delete Character & 2915 & 3438 & 332 & 59 & 94.20 & 94.60 & 98.31 & 94.62 \\
\cline{3-11}
& & Insert Character & 3049 & 3334 & 198 & 163 & 94.65 & 94.66 & 95.34 & 94.86 \\
\cline{3-11}
& & Replace Character & 3060 & 3408 & 187 & 89 & 95.91 & 95.99 & 97.45 & 96.11 \\
\tabucline[1.5pt]{-}
\multirow{9}{*}{Attention}
& - & Attack Free & 3213 & 3469 & 34 & 28 & 99.08 & 99.08 & 99.20 & 99.11 \\
\tabucline[1.5pt]{2-11}
& \multirow{4}{*}{Word-Level}
& Out Of Vocab & 2640 & 3462 & 607 & 35 & 90.48 & 91.89 & 99.00 & 91.51 \\
\cline{3-11}
& & Word Deletion & 3018 & 3446 & 229 & 51 & 95.85 & 96.05 & 98.54 & 96.10 \\
\cline{3-11}
& & Synonym Replacement & 2958 & 3420 & 289 & 77 & 94.57 & 94.84 & 97.80 & 94.92 \\
\cline{3-11}
& & Antonym Replacement & 3188 & 3458 & 59 & 39 & 98.55 & 98.56 & 98.88 & 98.60 \\
\tabucline[1.5pt]{2-11}
& \multirow{4}{*}{Character-Level} 
& Swap Letters & 3028 & 3450 & 219 & 47 & 96.06 & 96.25 & 98.66 & 96.29 \\
\cline{3-11}
& & Delete Character & 2853 & 3453 & 394 & 44 & 93.51 & 94.12 & 98.74 & 94.04 \\
\cline{3-11}
& & Insert Character & 2988 & 3448 & 259 & 49 & 95.43 & 95.70 & 98.60 & 95.72 \\
\cline{3-11}
& & Replace Character & 3045 & 3454 & 202 & 43 & 96.37 & 96.54 & 98.77 & 96.57 \\
\tabucline[1.5pt]{-}
\multirow{9}{*}{Transformer}
& - & Attack Free & 3165 & 3388 & 82 & 109 & 97.17 & 97.15 & 96.88 & 97.26 \\
\tabucline[1.5pt]{2-11}
& \multirow{4}{*}{Word-Level}
& Out Of Vocab & 2718 & 3254 &  529 &  243 & 88.55 & 88.90 & 93.05 & 89.40 \\
\cline{3-11}
& & Word Deletion & 2806 & 3301 & 441 & 196 & 90.55 & 90.84 & 94.40 & 91.20 \\
\cline{3-11}
& & Synonym Replacement & 3140 & 3343 & 107 & 154 & 96.13 & 96.11 & 95.60 & 96.24 \\
\cline{3-11}
& & Antonym Replacement & 3095 & 3421 & 152 & 76 & 96.62 & 96.67 & 97.83 & 96.78 \\
\tabucline[1.5pt]{2-11}
& \multirow{4}{*}{Character-Level} 
& Swap Letters & 2959 & 3139 & 288 & 358 & 90.42 & 90.40 & 89.76 & 90.67 \\
\cline{3-11}
& & Delete Character & 2527 & 3316 & 720 & 181 & 86.64 & 87.74 & 94.82 & 88.04 \\
\cline{3-11}
& & Insert Character & 2770 & 3275 & 477 & 222 & 89.64 & 89.93 & 93.65 & 90.36 \\
\cline{3-11}
& & Replace Character & 2889 & 3271 & 358 & 226 & 91.34 & 91.44 & 93.54 & 91.80 \\
\tabucline[1.5pt]{-}
\multirow{9}{*}{DistilBERT}
& - & Attack Free & 3085 & 3478 & 162 & 19 & 97.31 & 97.46 & 99.45 & 97.46 \\
\cline{2-11}
& \multirow{4}{*}{Word-Level}
& Out Of Vocab & 2964 & 3130 & 283 & 367 & 90.36 & 90.35 & 89.51 & 90.59 \\
\cline{3-11}
& & Word Deletion & 2571 & 3361 & 676 & 136 & 87.96 & 89.12 & 96.11 & 89.22 \\
\cline{3-11}
& & Synonym Replacement & 2974 & 3323 & 273 & 174 & 93.37 & 93.44 & 95.02 & 93.70 \\
\cline{3-11}
& & Antonym Replacement & 3138 & 3414 & 109 & 83 & 97.15 & 97.16 & 97.63 & 97.26 \\
\cline{2-11}
& \multirow{4}{*}{Character-Level} 
& Swap Letters & 2472 & 3315 & 775 & 182 & 85.81 & 87.10 & 94.80 & 87.39 \\
\cline{3-11}
& & Delete Character & 2750 & 3266 & 497 & 231 & 89.21 & 89.52 & 93.39 & 89.97 \\
\cline{3-11}
& & Insert Character & 2164 & 3318 & 1083 & 179 & 81.29 & 83.88 & 94.88 & 84.02 \\
\cline{3-11}
& & Replace Character & 2524 & 3326 & 723 & 171 & 86.74 & 87.90 & 95.11 & 88.15 \\
\tabucline[2pt]{-}
\end{tabu}}
\label{tab_r1Enron}
\end{table*}

\subsubsection{Sentence-Level Attack Results}
The results obtained using the spam weights scoring function for sentence-level attacks on the Enron Spam dataset \cite{d2} are presented in Table \ref{tab_swSentenceEnron}. The attention model proves to be more robust, benefiting from the effectiveness of its attention layer in handling long sentences. However, the CNN, dense, transformer, and distilBERT models show less resilience to these attacks. A common characteristic among these models is the absence of an LSTM layer, unlike traditional neural networks, which can process data with time steps of varying lengths. All models exhibit high resilience to add-ham sentence attacks but are significantly affected by add-spam sentence and add ham-spam sentence attacks.

With an add-ham sentence attack, the results indicate a slight increase in false positives, suggesting that some spam emails are misclassified as ham, resulting in a minimal decrease in accuracy. On the other hand, an add-spam sentence attack leads to a notable increase in false negatives and a substantial decrease in accuracy. Finally, with an add-ham-spam sentence attack, there is a rise in both false negatives and false positives, along with a significant decrease in accuracy.

\begin{table*}[!t]
\centering
\caption{Attack Results for the Enron Spam Dataset with Spam Weights at the Sentence-Level}
\resizebox{\textwidth}{!}{\begin{tabu}{c|c c c c c c c c c c}
\tabucline[2pt]{-}
\textbf{Model} & \textbf{Attack Level} & \textbf{Attack} & \textbf{TP} & \textbf{TN} & \textbf{FP} & \textbf{FN} & \textbf{Accuracy} & \textbf{Precision} & \textbf{Recall} & \textbf{F1 Score}\rule[-2ex]{0pt}{6ex}\\
\tabucline[2pt]{-}
\multirow{4}{*}{LSTM}
& - & Attack Free & 3208 & 3439 & 39 & 58 & 98.56 & 98.55 & 98.34 & 98.61 \\
\tabucline[1.5pt]{2-11}
& \multirow{3}{*}{Sentences-Level} 
& Add Ham Sentence & 3157 & 3481 &  90 & 16 & 98.43 & 98.49 & 99.54 & 98.50 \\
\cline{3-11}
&  & Add Spam Sentence & 3233 & 2594 & 14 &  903 & 86.40 & 88.82 & 74.18 & 84.98 \\
\cline{3-11}
&  & Add Ham-Spam Sentence & 3157 & 2594 & 90 & 903 & 85.28 & 87.20 & 74.18 & 83.93 \\ 
\tabucline[1.5pt]{-}
\multirow{4}{*}{Dense}
& - & Attack Free & 3210 & 3426 & 37 & 71 & 98.40 & 98.38 & 97.97 & 98.45 \\
\tabucline[1.5pt]{2-11}
& \multirow{3}{*}{Sentences-Level} 
& Add Ham Sentence &2653 & 3496 & 594 &  1 & 91.18 & 92.72 & 99.97 & 92.16 \\
\cline{3-11}
&  & Add Spam Sentence & 3246 & 1422 &  1 & 2075 & 69.22 & 80.47 & 40.66 & 57.80 \\
\cline{3-11}
&  & Add Ham-Spam Sentence & 2653 & 1422 & 594 & 2075 & 60.42 & 63.32 & 40.66 & 51.59 \\ 
\tabucline[1.5pt]{-}
\multirow{4}{*}{CNN}
& - & Attack Free & 3207 & 3463 & 40 & 34 & 98.90 & 98.90 & 99.03 & 98.94 \\
\tabucline[1.5pt]{2-11}
& \multirow{3}{*}{Sentence-Level} 
& Add Ham Sentence & 3247 &  755 &  0 & 2742 & 59.34 & 77.11 & 21.59 & 35.51 \\
\cline{3-11}
&  & Add Spam Sentence & 3231 & 2936 & 16 &  561 & 91.44 & 92.33 & 83.96 & 91.05 \\
\cline{3-11}
&  & Add Ham-Spam Sentence & 3144 & 755 & 103 & 2742 & 57.81 & 70.71 & 21.59 & 34.67 \\ 
\tabucline[1.5pt]{-}
\multirow{4}{*}{Attention}
& - & Attack Free & 3213 & 3469 & 34 & 28 & 99.08 & 99.08 & 99.20 & 99.11 \\
\tabucline[1.5pt]{2-11}
& \multirow{3}{*}{Sentence-Level} 
& Add Ham Sentence & 2973 & 3492 & 274 & 5 & 95.86 & 96.28 & 99.86 & 96.16 \\
\cline{3-11}
&  & Add Spam Sentence & 2973 & 3492 & 274 & 5 & 95.86 & 96.28 & 99.86 & 96.16 \\
\cline{3-11}
&  & Add Ham-Spam Sentence & 2973 & 2936 & 274 & 561 & 87.62 & 87.79 & 83.96 & 87.55 \\
\tabucline[1.5pt]{-}
\multirow4{*}{Transformer}
& - & Attack Free & 3165 & 3388 & 82 & 109 & 97.17 & 97.15 & 96.88 & 97.26 \\
\tabucline[1.5pt]{2-11}
& \multirow{3}{*}{Sentence-Level} 
& Add Ham Sentence & 3083 &  3409 & 164 & 88 & 96.26 & 96.32 & 97.48 & 96.44 \\
\cline{3-11}
&  & Add Spam Sentence & 3243 &  536 &  4 & 2961 & 56.03 & 75.77 & 15.33 & 26.55 \\
\cline{3-11}
&  & Add Ham-Spam Sentence & 3083 &  536 & 164 & 2961 & 53.66 & 63.79 & 15.33 & 25.54 \\
\tabucline[1.5pt]{-}
\multirow4{*}{DistilBERT}
& - & Attack Free & 3085 & 3478 & 162 & 19 & 97.31 & 97.46 & 99.45 & 97.46 \\
\tabucline[1.5pt]{2-11}
& \multirow{3}{*}{Sentence-Level} 
& Add Ham Sentence & 2728 & 3490 & 519 & 7 & 92.20 & 93.40 & 99.80 & 92.99 \\
\cline{3-11}
&  & Add Spam Sentence & 3234 & 1905 & 13 & 1592 & 76.20 & 83.17 & 54.48 & 70.36 \\
\cline{3-11}
&  & Add Ham-Spam Sentence & 3246 & 1422 & 1 & 2075 & 69.22 & 80.47 & 40.66 & 57.80\\
\tabucline[2pt]{-}
\end{tabu}}
\label{tab_swSentenceEnron}
\end{table*}

\subsubsection{Paragraph-Level Attack Results}
Paragraph-level attacks consisting of 500 spam and 500 non-spam emails were generated using AI. The email subjects were randomly assigned and covered various topics. These AI-generated attacks were then tested on pre-trained models that had been trained on the Enron Spam dataset \cite{d2}. The results are presented in Table \ref{tab_aiResults}.

The performance of deep learning models varies significantly across metrics such as accuracy, precision, recall, and F1 score, largely influenced by their architectures and capabilities. DistilBERT consistently outperforms other models, particularly in precision, recall, and F1 score, due to its transformer-based architecture, which effectively captures long-range dependencies and contextual information. This advantage is especially critical for paragraph-level tasks, where relationships span across words and sentences. DistilBERT's ability to process entire sequences of words simultaneously, coupled with pre-training on extensive corpora, enables it to identify subtle patterns and achieve high accuracy in tasks like distinguishing spam from non-spam emails. Its capability to handle imbalanced datasets further enhances both precision and recall, contributing to an improved F1 score. Unlike older models such as LSTMs or CNNs, which rely on sequential or convolutional operations, DistilBERT processes text bidirectionally. This allows it to consider the context of a word both before and after its position in a sentence, leading to more accurate predictions. Pre-trained transformer models like DistilBERT generalize effectively across various natural language processing tasks, making them more robust compared to task-specific models such as LSTMs or dense networks.

On the other hand, the LSTM model follows with decent accuracy, but its recall scores are relatively low because, while it is designed to handle sequential data well, it struggles with long-term dependencies and tends to miss some positive cases (spam). This limitation likely contributes to its lower recall, as it cannot capture complex patterns as effectively as transformer-based models like distilBERT. Transformer and CNN perform similarly, with moderate accuracies. While both architectures are capable of identifying certain patterns in the data, their low recall and moderate precision suggest that they are not as effective at capturing the nuanced differences between spam and non-spam emails. However, the transformer model does not have the benefits of pre-training on large datasets. The CNN, which is commonly used for image recognition, may not be well-suited for text-based tasks, as it primarily focuses on local patterns rather than long-range dependencies. In addition, the dense and attention models show the poorest performance, with low accuracy, precision, recall, and F1 scores. The dense model, being a fully connected feed-forward network, lacks the ability to effectively handle sequential dependencies in text data. The attention model, despite its focus mechanism, likely underperformed due to the limited complexity of its architecture when compared to more advanced models like transformer. Overall, the superior architecture of distilBERT and its pre-training enable it to outperform the others, particularly in recall, where it captures a much higher proportion of true positives, while the other models struggle with both recall and precision.

\begin{table*}[!t]
\caption{Attack Results for AI-Generated Dataset at the Paragraph-Level}
\begin{center}
\resizebox{\columnwidth}{!}{\begin{tabu}{c|c c c c c c c c c}
\hline
\textbf{Model} & \textbf{Attack Level} & \textbf{TP} & \textbf{TN} & \textbf{FP} & \textbf{FN} & \textbf{Accuracy} & \textbf{Precision} & \textbf{Recall} & \textbf{F1 Score} \rule[-2ex]{0pt}{6ex}\\
\hline
\multirow{1}{*}{LSTM} 
& Paragraph-Level & 428 & 28 & 72 & 472 & 45.60 & 37.78 & 5.60 & 9.33\\
\hline
\multirow{1}{*}{Dense} 
& Paragraph-Level & 287 & 35 & 213 & 465 & 32.20 & 26.14 & 7.00 & 9.36 \\
\hline
\multirow{1}{*}{CNN} 
& Paragraph-Level & 360 & 35 & 140 & 465 & 39.50 & 31.82 & 7.00 & 10.37 \\
\hline
\multirow{1}{*}{Attention} 
& Paragraph-Level & 370 & 28 & 130 & 472 & 39.80 & 30.83 & 5.60 & 8.51 \\
\hline
\multirow{1}{*}{Transformer} 
& Paragraph-Level & 358 & 38 & 142 & 462 & 39.60 & 32.38 & 7.60 & 11.18 \\
\hline
\multirow{1}{*}{DistilBERT} 
& Paragraph-Level & 220 & 490 & 280 & 10 & 71.00 & 79.64 & 98.00 & 77.17 \\
\hline
\end{tabu}}
\label{tab_aiResults}
\end{center}
\end{table*}

\section{General Discussion}

A discussion on some of the interesting findings that found while studying different spam filters and adversarial attacks will be presented. Additionally, the challenges in this field will be highlighted.

While all models perform well in attack-free scenarios, their robustness varies under different types of adversarial attacks, highlighting the need for improved defenses against such attacks. Comparing the performance of the filters at the character, word, and sentence levels, the LSTM model shows the most significant decrease in accuracy for the spam weights scoring function at all levels (character-level, word-level, and sentence-level) due to the calculation of spam weights using an LSTM model. On the contrary, the dense, transformer, and distilBERT filters show the largest decrease in accuracy, while the attention filter shows the smallest decrease for the R1S and attention weights scoring functions. Consequently, the dense, transformer, and distilBERT models are not robust against character, word, and sentence-level attacks compared to others. In contrast, the attention model uses the attention layer, which plays an important role in improving the performance and interpretability of NLP models by enabling them to focus on relevant information. Incorporating both attention and LSTM layers enhances the filter's resistance to attacks. Conversely, when evaluating the performance of the models on AI-generated paragraph-level tasks, distilBERT significantly outperforms the others, making it the most effective model for this task due to knowledge distillation and its better representations. DistilBERT is more robust due to knowledge distillation and its better representations, whereas the transformer model is vulnerable at the paragraph-level.

When comparing the different levels of attack, it can be seen that certain word-level attacks (OOV and word deletion) cause a more pronounced drop in accuracy than character-level attacks, while others (synonym and antonym replacement) cause almost no drop in accuracy across all models. The models show the highest resilience to synonym replacement attack, which maintains the semantic integrity of the original email with only a minor drop in performance, whereas OOV has the most detrimental effect across all models, with accuracy dropping significantly. However, the attack rates for the character-level attacks are close together, and the most significant decrease in accuracy was observed for the character insertion and deletion attacks. Some models have shown that these attacks can reduce accuracy more than word-level attacks. Among sentence-level attacks, adding ham and spam sentences reduced accuracy more in the CNN, dense, and transformer models than in other levels. Overall, paragraph-level attacks are found to cause a more substantial decrease across all metrics.

It is also remarkable to compare the proposed scoring functions. Furthermore, the results obtained with the R1S scoring function are similar to those obtained with the attention weights scoring function, with a slight decrease in accuracy observed for word-level and character-level attacks. Although both scoring functions produce similar results, the attention weights scoring function performs better than the R1S scoring function in terms of performance because the R1S scoring function processes the words in the corpus one by one. There is a more pronounced decrease in the effectiveness of spam filters when using the spam weights scoring function compared to the attention weights and R1S scoring functions. The spam weights technique efficiently identifies words by estimating their spam probabilities. Compared to other scoring functions, spam weights have shown superior effectiveness in combating spam filters, particularly in terms of speed and reducing the success rate of the filters.

The results are explained here for the Enron Spam dataset \cite{d2}. The attacks were also applied to other datasets, and results were obtained. The results for the SpamAssassin dataset \cite{d1} and the TREC2007 dataset \cite{d3} are provided in the supplementary material \cite{supp}, showing the performance with spam weights, attention weights, and R1S scoring functions. In the SpamAssassin dataset \cite{d1}, the attacks were applied to 3\% of the corpus size, as in the Enron Spam dataset. Spam weights showed the most significant decrease in all filters, while attention weights and R1S showed a decrease in accuracy in some filters. Similar results were obtained with the same attacks as in the Enron Spam dataset. However, the transformer and distilBERT models are more robust on the SpamAssassin \cite{d1} dataset than on the Enron Spam dataset \cite{d2} because the emails in the Enron Spam dataset \cite{d2} have larger message sizes. For the TREC2007 dataset \cite{d3}, the corpus size is almost 10 times larger than the other datasets, so attacks were applied to 0.3\% of the words in this dataset. The results are similar to those from the Enron Spam dataset \cite{d2}, but since it is the dataset with the largest message size, the attacks took longer to implement, and their success was lower compared to other datasets. Better success will likely be achieved when the percentage of words affected increases. The results of sentence-level attacks are also provided in the supplementary material \cite{supp} for the SpamAssassin \cite{d1} and TREC2007 datasets \cite{d3}.

When comparing the mentioned spam datasets to the results from AI-generated paragraph-level data, it becomes increasingly clear that spam filters often struggle to detect the latter effectively. This vulnerability raises significant concerns, particularly given that the spam datasets, such as the Enron Spam dataset \cite{d2}, the SpamAssassin dataset \cite{d1}, and the TREC 2007 dataset \cite{d3}, are relatively outdated. These older datasets may not accurately reflect the evolving tactics employed by modern AI-generated spam, which can utilize sophisticated language patterns and personalization techniques that traditional filters might not recognize. Additionally, these established datasets contain a larger volume of data compared to AI-generated dataset, which can create an imbalance in the training process for spam detection models. The effectiveness of spam filters heavily relies on their training data; thus, utilizing obsolete datasets can lead to significant gaps in performance. This discrepancy can result in higher rates of false negatives, where genuine spam slips through the filters, and false positives, where legitimate emails are incorrectly flagged as spam.

In the study, it is noticed several challenges in this field. The concept of imperceptibility poses a significant limitation in textual data because, unlike continuous image data, text data is discrete. In the image domain, perturbations can be nearly imperceptible to humans perception, causing discrepancies with models. In contrast, in the text domain, even small changes are usually noticeable to humans and can significantly alter the sentence's meaning.

The prepared adversarial email can result in a large number of changed words. For example, if the email designed to bypass the spam filter in the experimental dataset contains too many words selected by scoring functions, the number of targeted words may be high. This may not be practical to implement in the real world. If the original email is long, it is relatively easy to hide the changes made to it when the attack is applied. Additionally, when the attacks are implemented, it is difficult to verify whether the resulting email is still a spam email or a raw email. These challenges highlight the complexities involved in creating and defending against adversarial attacks on spam filters, emphasizing the need for robust, adaptable, and ethical approaches in both offensive and defensive strategies.

\section{Conclusion}
This comprehensive analysis examines various attack vectors at the word, character, sentence level, and AI-generated paragraph-level and presents a number of attack strategies targeting deep learning models used in spam classification. Despite significant progress in the area of adversarial learning, particularly in image recognition, the area of text adversarial attacks remains relatively unexplored and represents a promising area for further research and innovation. This study aims to address this gap by analyzing adversarial attacks against six prominent deep learning-based spam filters. 

Furthermore, this study introduces a novel scoring function, known as spam weights, which is designed to intelligently identify which segments of text are most amenable to manipulation to achieve adversarial goals. The attention weights scoring function is also explored for adversarial attacks against spam filters for the first time in this study. What sets spam weights scoring function apart is its ability to deliver results comparable to established scoring functions such as attention weights and R1S, but with a significantly reduced computational overhead. This efficiency not only streamlines the adversarial generation process, but also improves scalability, facilitating the creation of diverse attack types across a variety of deep learning models and datasets. This study also investigates sentence-level and AI-generated paragraph-level attacks, for the first time, against NLP-based systems. 

Through careful experimentation and evaluation across six different models and three real-world spam email datasets, the results highlight the effectiveness of spam weights in identifying the most effective words for manipulation, providing invaluable insights into the dynamics of adversarial attacks in the field of text classification. This claim is corroborated by implementing attacks at four different levels: word, character, sentence, and AI-generated paragraph-level. By shedding light on these effective strategies for perturbing textual data, the study lays a solid foundation for the development of robust defences against adversarial attacks in spam filtering.

The performance of the AI-generated paragraph-level demonstrates its effectiveness in evaluating model accuracy, highlighting distinct differences in how well various deep learning models can classify the data. The distilBERT outperforms the other models by a wide margin, particularly in precision, recall, and F1 score, indicating that it is the most effective for this task, while the dense and attention models have the poorest performance, with low accuracy, precision, recall, and F1 scores, making them less suitable for the classification task.

\appendix


 \bibliographystyle{apalike} 
 \bibliography{references}





\end{document}


\begin{frontmatter}



\title{Supplementary Material for: A Comprehensive Analysis of Adversarial Attacks against Spam Filters}


\author[inst1]{Esra Hotoğlu}

\affiliation[inst1]{organization={WISE Lab., Department of Computer Engineering},
            addressline={Hacettepe University}, 
            city={Ankara},
            country={Turkey}}

\author[inst1]{Sevil Sen\corref{cor1}}
\ead{ssen@cs.hacettepe.edu.tr}
\cortext[cor1]{Corresponding author}

\author[inst2]{Burcu Can}

\affiliation[inst2]{organization={Department of Computing Science and Mathematics},
            addressline={University of Stirling}, 
            city={Stirling},
            country={UK}}

\end{frontmatter}


\section{Supplementary Material Introduction}

This supplementary material provides additional tables that support the findings presented in the main manuscript titled "A Comprehensive Analysis of Adversarial Attacks against Spam Filters." These tables present the results for the SpamAssassin and TREC2007 datasets, offering further insights into the findings.

Table~\ref{tab_SpamAssassin} and Table~\ref{tab_trec} show the baseline performance of the classifiers for the SpamAssassin and TREC2007 datasets, respectively. Tables~\ref{tab_swSpamAssassin}, \ref{tab_awSpamAssassin}, and \ref{tab_r1SpamAssassin} display the attack results for the SpamAssassin dataset at both character and word levels. Similarly, Tables~\ref{tab_swTrec}, \ref{tab_awTrec}, and \ref{tab_r1Trec} present the attack results for the TREC2007 dataset at these levels. Finally, Tables~\ref{tab_swSentenceSpamAssassin} and \ref{tab_swSentenceTrec} provide the sentence-level attack results for both datasets. Together, these supplementary materials offer a deeper understanding of the results discussed in the main manuscript.

\begin{table}[htbp]
\caption{Attack-Free Results for the SpamAssassin Dataset}
\begin{center}
\small
\resizebox{\textwidth}{!}{\begin{tabular}{c c c c c c c c c}
\hline
\textbf{Model} & \textbf{TP} & \textbf{TN} & \textbf{FP} & \textbf{FN} & \textbf{Accuracy} & \textbf{Precision} & \textbf{Recall} & \textbf{F1 Score} \\
\hline
LSTM & 1393 & 468 & 1 & 9 & 99.46 & 99.57 & 98.11 & 98.94 \\
\hline
Dense & 1391 & 464 & 3 & 13 & 99.14 & 99.21  & 97.27 & 98.30 \\
\hline
CNN & 1387 & 475 & 7 & 2 & 99.51 & 99.20 & 99.58 & 99.06 \\
\hline
Attention & 1389 & 471 & 5 & 6 & 99.41 & 99.25 & 98.74 & 98.84 \\
\hline
Transformer & 1389 & 460 & 5 & 17 & 98.82 & 98.86 & 96.44 & 97.66 \\
\hline
DistilBERT & 1382 & 458 & 12 & 19 & 98.34 & 98.04 & 96.01 & 96.72 \\
\hline
\end{tabular}}
\label{tab_SpamAssassin}
\end{center}
\end{table}

\begin{table}[htbp]
\caption{Attack-Free Results for the TREC2007 Dataset}
\begin{center}
\resizebox{\textwidth}{!}{\begin{tabular}{c c c c c c c c c}
\hline
\textbf{Model} & \textbf{TP} & \textbf{TN} & \textbf{FP} & \textbf{FN} & \textbf{Accuracy} & \textbf{Precision} & \textbf{Recall} & \textbf{F1 Score} \\
\hline
LSTM & 5035 & 9958 & 38 & 53 & 99.40 & 99.29 & 99.47 & 99.55 \\
\hline
Dense & 5036 & 9973 & 37 & 38 & 99.50 & 99.44 & 99.62 & 99.63 \\
\hline
CNN & 5034 & 9987 & 39 & 24 & 99.58 & 99.57 & 99.76 & 99.69 \\
\hline
Attention & 5037 & 9997 & 36 & 14 & 99.67 & 99.68 & 99.86 & 99.75 \\
\hline
Transformer & 4895 & 9963 & 178 & 48 & 98.50 & 98.64 & 99.52 & 98.88 \\
\hline
DistilBERT & 5061 & 10001 & 12 & 10 & 99.85 & 99.84 & 99.90 & 99.89 \\
\hline
\end{tabular}}
\label{tab_trec}
\end{center}
\end{table}

\begin{table*}[!t]
\centering
\caption{Attack Results for SpamAssassin Dataset with Spam Weights}
\resizebox{\textwidth}{!}{\begin{tabular}{c|c c c c c c c c c c}
\hline\hline
\textbf{Model} & \textbf{Attack Level} & \textbf{Attack} & \textbf{TP} & \textbf{TN} & \textbf{FP} & \textbf{FN} & \textbf{Accuracy} & \textbf{Precision} & \textbf{Recall} & \textbf{F1 Score}\rule[-2ex]{0pt}{6ex}\\
\hline\hline
\multirow{9}{*}{LSTM}
& - & Attack Free & 1393 & 468 & 1 & 9 & 99.46 & 99.57 & 98.11 & 98.94  \\
\cline{2-11}
& \multirow{4}{*}{Word-Level} 
& Out Of Vocab & 121 & 477 & 1273 & 0 & 31.96 & 63.63 & 100.00 & 42.84  \\
\cline{3-11}
& & Word Deletion & 121 & 477 & 1273 & 0 & 31.96 & 63.63 & 100.00 & 42.84  \\
\cline{3-11}
& & Synonym Replacement & 1370 & 444 & 24 & 33 & 96.95 & 96.26 & 93.08 & 93.97  \\
\cline{3-11}
& & Antonym Replacement & 1391 & 450 & 3 & 27 & 98.40 & 98.72 & 94.34 & 96.77 \\
\cline{2-11}
& \multirow{4}{*}{Character-Level} 
& Swap Letters & 536  & 475 & 858 & 2 & 54.04 & 67.63 & 99.58 & 52.49  \\
\cline{3-11}
& & Delete Character & 530 & 475 & 864 & 2 & 53.71 & 67.55 & 99.58 & 52.31  \\
\cline{3-11}
& & Insert Character & 124 & 477 & 1270 & 0 & 32.12 & 63.65 & 100.00 & 42.90  \\
\cline{3-11}
& & Replace Character & 545 & 475 & 849 & 2 & 54.52 & 67.76 & 99.58 & 52.75  \\
\hline\hline
\multirow{9}{*}{Dense}
& - & Attack Free & 1391 & 464 & 3 & 13 & 99.14 & 99.21 & 97.27 & 98.30  \\
\cline{2-11}
& \multirow{4}{*}{Word-Level} 
& Out Of Vocab & 244 & 471 & 1150 & 6 & 38.21 & 63.33 & 98.74 & 44.90  \\
\cline{3-11}
& & Word Deletion & 243 & 471 & 1151 & 6 & 38.16 & 63.31 & 98.74 & 44.88  \\
\cline{3-11}
& & Synonym Replacement & 1375 & 463 & 19 & 14 & 98.24 & 97.53 & 97.06 & 96.56  \\
\cline{3-11}
& & Antonym Replacement & 1385 & 465 &  9 & 12 & 98.88 & 98.62 & 97.48 & 97.79  \\
\cline{2-11}
& \multirow{4}{*}{Character-Level} 
& Swap Letters & 339  & 471  & 1055 & 6 &  43.29 & 64.56 & 98.74 &  47.03  \\
\cline{3-11}
& & Delete Character & 329 & 474 & 1065 & 3 & 42.92 & 64.95 & 99.37 & 47.02  \\
\cline{3-11}
& & Insert Character & 234 & 471 & 1160 & 6 & 37.68 & 63.19 & 98.74 & 44.69  \\
\cline{3-11}
& & Replace Character & 322 & 472 & 1072 & 5 & 42.44 & 64.52 & 98.95 & 46.71  \\
\hline\hline
\multirow{9}{*}{CNN}
& - & Attack Free & 1387 & 475 & 7 & 2 & 99.51 & 99.20 & 99.58 & 99.06  \\
\cline{2-11}
& \multirow{4}{*}{Word-Level} 
& Out Of Vocab & 433 & 472 &  961 & 5 & 48.37 & 65.90 &  98.95 & 49.42  \\
\cline{3-11}
& & Word Deletion & 433 & 472 &  961 & 5 & 48.37 & 65.90 &  98.95 & 49.42  \\
\cline{3-11}
& & Synonym Replacement & 1378 & 458 & 16 & 19 & 98.13 & 97.63 & 96.02 & 96.32  \\
\cline{3-11}
& & Antonym Replacement & 1385 & 459 & 9 & 18 & 98.56 & 98.40 & 96.23 & 97.14  \\
\cline{2-11}
& \multirow{4}{*}{Character-Level} 
& Swap Letters & 729  & 467  &  665 &  10 &  63.92 & 69.95 & 97.90 & 58.05  \\
\cline{3-11}
& & Delete Character & 689 & 468 & 705 & 9 & 61.84 & 69.30 & 98.11 & 56.73  \\
\cline{3-11}
& & Insert Character & 429 & 473 & 965 & 4 & 48.21 & 65.98 & 99.16 & 49.40  \\
\cline{3-11}
& & Replace Character & 721 & 467 & 673 & 10 & 63.50 & 69.80 & 97.90 & 57.76  \\
\hline\hline
\multirow{9}{*}{Attention}
& - & Attack Free & 1389 & 471 & 5 & 6 & 99.41 & 99.25 & 98.74 & 98.84 \\
\cline{2-11}
& \multirow{4}{*}{Word-Level}
& Out Of Vocab & 450 & 477 & 944 & 0 & 49.55 & 66.78 & 100.00 & 50.26  \\
\cline{3-11}
& & Word Deletion & 450 & 477 & 944 & 0 & 49.55 & 66.78 & 100.00 & 50.26  \\
\cline{3-11}
& & Synonym Replacement & 1387 & 471 & 7 & 6 & 99.31 & 99.05 & 98.74 & 98.64  \\
\cline{3-11}
& & Antonym Replacement & 1387 & 472 & 7 & 5 & 99.36 & 99.09 & 98.95 & 98.74  \\
\cline{2-11}
& \multirow{4}{*}{Character-Level} 
& Swap Letters & 775 & 476 & 619 & 1 & 66.86 & 71.67 & 99.79 & 60.56  \\
\cline{3-11}
& & Delete Character & 724 & 476 & 670 & 1 & 64.14 & 70.70 & 99.79 & 58.66  \\
\cline{3-11}
& & Insert Character & 443 & 477 & 951 & 0 & 49.17 & 66.70 & 100.00 & 50.08  \\
\cline{3-11}
& & Replace Character & 772 & 476 & 622 & 1 & 66.70 & 71.61 & 99.79 & 60.44  \\
\hline\hline
\multirow{9}{*}{Transformer}
& - & Attack Free & 1389 & 460 & 5 & 17 & 98.82 & 98.86 & 96.44 & 97.66  \\
\cline{2-11}
& \multirow{4}{*}{Word-Level}
& Out Of Vocab &648 & 475 &  746 & 2 & 60.02 & 69.30 &  99.58 & 55.95  \\
\cline{3-11}
& & Word Deletion & 648 & 475 &  746 & 2 & 60.02 & 69.30 &  99.58 & 55.95  \\
\cline{3-11}
& & Synonym Replacement & 1374 & 466 & 20 & 11 & 98.34 & 97.55 & 97.69 & 96.78  \\
\cline{3-11}
& & Antonym Replacement & 1389 & 461 &  5 & 16 & 98.88 & 98.89 & 96.65 & 97.77  \\
\cline{2-11}
& \multirow{4}{*}{Character-Level} 
& Swap Letters & 960 & 471 &  434 &  6 & 76.48 & 75.71 & 98.74 & 68.16  \\
\cline{3-11}
& & Delete Character & 923 & 471 &  471 & 6 & 74.51 & 74.68 & 98.74 & 66.38  \\
\cline{3-11}
& & Insert Character & 636 & 474 &  758 & 3 & 59.33 & 69.00 &  99.37 & 55.47  \\
\cline{3-11}
& & Replace Character & 942 & 470 &  452 &  7 & 75.47 & 75.12 & 98.53 & 67.19  \\
\hline\hline
\multirow{9}{*}{DistilBERT}
& - & Attack Free & 1382 & 458 & 12 & 19 & 98.34 & 98.04 & 96.01 & 96.72 \\
\cline{2-11}
& \multirow{4}{*}{Word-Level}
& Out Of Vocab & 1374 & 439 & 20 & 38 & 96.90 & 96.47 & 92.03 & 93.80 \\
\cline{3-11}
& & Word Deletion & 1305 & 465 & 89 & 12 & 94.60 & 91.51 & 97.48 & 90.20 \\
\cline{3-11}
& & Synonym Replacement & 1384 & 450 & 10 & 27 & 98.01 & 97.95 & 94.33 & 96.05 \\
\cline{3-11}
& & Antonym Replacement & 1383 & 455 & 11 & 22 & 98.23 & 98.03 & 95.38 & 96.50 \\
\cline{2-11}
& \multirow{4}{*}{Character-Level} 
& Swap Letters & 1352 & 436 & 42 & 41 & 95.56 & 94.13 & 91.40 & 91.30 \\
\cline{3-11}
& & Delete Character & 1370 & 443 & 24 & 34 & 96.90 & 96.21 & 92.87 & 93.85 \\
\cline{3-11}
& & Insert Character & 1204 & 447 & 190 & 30 & 88.24 & 83.87 & 93.71 & 80.25 \\
\cline{3-11}
& & Replace Character & 1310 & 454 & 84 & 23 & 94.28 & 91.330 & 95.17& 89.45 \\
\hline\hline
\end{tabular}}
\label{tab_swSpamAssassin}
\end{table*}

\begin{table*}[!t]
\centering
\caption{Attack Results for SpamAssassin Dataset with Attention Weights}
\resizebox{\textwidth}{!}{\begin{tabular}{c|c c c c c c c c c c}
\hline\hline
\textbf{Model} & \textbf{Attack Level} & \textbf{Attack} & \textbf{TP} & \textbf{TN} & \textbf{FP} & \textbf{FN} & \textbf{Accuracy} & \textbf{Precision} & \textbf{Recall} & \textbf{F1 Score}\rule[-2ex]{0pt}{6ex}\\
\hline\hline
\multirow{9}{*}{LSTM}
& - & Attack Free & 1393 & 468 & 1 & 9 & 99.46 & 99.57 & 98.11 & 98.94  \\
\cline{2-11}
& \multirow{4}{*}{Word-Level} 
& Out Of Vocab & 1309 & 275 &  85 & 202 & 84.66 & 81.51 & 57.65 & 65.71  \\
\cline{3-11}
& & Word Deletion & 1310 & 278 &  84 & 199 & 84.87 & 81.80 & 58.28 & 66.27  \\
\cline{3-11}
& & Synonym Replacement & 1376 & 451 & 18 & 26 & 97.65 & 97.15 & 94.55 & 95.35  \\
\cline{3-11}
& & Antonym Replacement & 1389 & 453 & 5 & 24 & 98.45 & 98.60 & 94.97 & 96.90  \\
\cline{2-11}
& \multirow{4}{*}{Character-Level} 
& Swap Letters & 1253 & 372 & 141 & 105 & 86.85 & 82.39 & 77.99 & 75.15  \\
\cline{3-11}
& & Delete Character & 1220 & 359 & 174 & 118 & 84.39 & 79.27 & 75.26 & 71.09  \\
\cline{3-11}
& & Insert Character & 1290 & 281 & 104 & 196 & 83.97 & 79.90 & 58.91 & 65.20  \\
\cline{3-11}
& & Replace Character & 1251 & 362 & 143 & 115 & 86.21 & 81.63 & 75.89 & 73.73  \\
\hline\hline
\multirow{9}{*}{Dense}
& - & Attack Free & 1391 & 464 & 3 & 13 & 99.14 & 99.21 & 97.27 & 98.30  \\
\cline{2-11}
& \multirow{4}{*}{Word-Level} 
& Out Of Vocab & 390 & 435 & 1004 & 42 & 44.09 & 60.25 & 91.19 & 45.41  \\
\cline{3-11}
& & Word Deletion & 388 & 437 & 1006 &  40 & 44.09 & 60.47 & 91.61 & 45.52  \\
\cline{3-11}
& & Synonym Replacement & 1377 & 466 & 17 & 11 & 98.50 & 97.84 & 97.69 & 97.08  \\
\cline{3-11}
& & Antonym Replacement & 1385 & 459 & 9 & 18 & 98.56 & 98.40 & 96.23 & 97.14  \\
\cline{2-11}
& \multirow{4}{*}{Character-Level} 
& Swap Letters & 1025 & 414 & 369 & 63 & 76.91 & 73.54 & 86.79 & 65.71  \\
\cline{3-11}
& & Delete Character & 1061 & 407 & 333 & 70 & 78.46 & 74.41 & 85.32 & 66.89  \\
\cline{3-11}
& & Insert Character & 405 & 430 & 989 & 47 & 44.63 & 59.95 & 90.15 & 45.36  \\
\cline{3-11}
& & Replace Character & 1025 & 423 & 369 & 54 & 77.39 & 74.20 & 88.68 & 66.67  \\
\hline\hline
\multirow{9}{*}{CNN}
& - & Attack Free & 1387 & 475 & 7 & 2 & 99.51 & 99.20 & 99.58 & 99.06  \\
\cline{2-11}
& \multirow{4}{*}{Word-Level} 
& Out Of Vocab & 452 & 303 & 942 & 174 & 40.35 & 48.27 & 63.52 & 35.19  \\
\cline{3-11}
& & Word Deletion & 452 & 302 & 942 & 175 & 40.30 & 48.18 & 63.31 & 35.10  \\
\cline{3-11}
& & Synonym Replacement & 1388 & 445 & 6 & 32 & 97.97 & 98.21 & 93.29 & 95.91  \\
\cline{3-11}
& & Antonym Replacement & 1387 & 457 & 7 & 20 & 98.56 & 98.53 & 95.81 & 97.13  \\
\cline{2-11}
& \multirow{4}{*}{Character-Level} 
& Swap Letters & 1160 & 331 & 234 & 146 & 79.69 & 73.70 & 69.39 & 63.53  \\
\cline{3-11}
& & Delete Character & 1193 & 293 & 201 & 184 & 79.42 & 72.97 & 61.43 & 60.35  \\
\cline{3-11}
& & Insert Character &515 & 275 & 879 & 202 & 42.22 & 47.83 & 57.65 & 33.72  \\
\cline{3-11}
& & Replace Character & 1131 & 336 & 263 & 141 & 78.41 & 72.50 & 70.44 & 62.45  \\
\hline\hline
\multirow{9}{*}{Attention}
& - & Attack Free & 1389 & 471 & 5 & 6 & 99.41 & 99.25 & 98.74 & 98.84  \\
\cline{2-11}
& \multirow{4}{*}{Word-Level}
& Out Of Vocab & 1384 & 397 & 10 & 80 & 95.19 & 96.04 & 83.23 & 89.82  \\
\cline{3-11}
& & Word Deletion & 1384 & 396 & 10 & 81 & 95.14 & 96.00 & 83.02 & 89.69  \\
\cline{3-11}
& & Synonym Replacement & 1389 & 457 & 5 & 20 & 98.66 & 98.75 & 95.81 & 97.34  \\
\cline{3-11}
& & Antonym Replacement & 1387 & 473 & 7 & 4 & 99.41 & 99.13 & 99.16 & 98.85  \\
\cline{2-11}
& \multirow{4}{*}{Character-Level} 
& Swap Letters & 805 & 471 & 589 & 6 & 68.20 & 71.85 & 98.74 & 61.29  \\
\cline{3-11}
& & Delete Character & 848 & 468 & 546 & 9 & 70.34 & 72.55 & 98.11 & 62.78  \\
\cline{3-11}
& & Insert Character & 1381 & 392 & 13 & 85 & 94.76 & 95.50 & 82.18 & 88.89   \\
\cline{3-11}
& & Replace Character & 1203 & 467 & 191 &  10 & 89.26 & 85.07 & 97.90 & 82.29   \\
\hline\hline
\multirow{9}{*}{Transformer}
& - & Attack Free & 1389 & 460 & 5 & 17 & 98.82 & 98.86 & 96.44 & 97.66  \\
\cline{2-11}
& \multirow{4}{*}{Word-Level}
& Out Of Vocab & 1266 & 370 &  128 & 107 & 87.44 & 83.25 & 77.57 & 75.90  \\
\cline{3-11}
& & Word Deletion & 1265 & 371 &  129 & 106 & 87.44 & 83.23 & 77.78 & 75.95  \\
\cline{3-11}
& & Synonym Replacement & 1384 & 453 & 10 & 24 & 98.18 & 98.07 & 94.97 & 96.38  \\
\cline{3-11}
& & Antonym Replacement & 1389 & 463 & 5 & 14 & 98.98 & 98.97 & 97.06 & 97.99  \\
\cline{2-11}
& \multirow{4}{*}{Character-Level} 
& Swap Letters & 1305 & 416 &  89 & 61 & 91.98 & 88.96 & 87.21 & 84.73  \\
\cline{3-11}
& & Delete Character & 1308 & 398 & 86 & 79 & 91.18 & 88.27 & 83.44 & 82.83  \\
\cline{3-11}
& & Insert Character & 1257 & 369 & 137 & 108 & 86.91 & 82.51 & 77.36 & 75.08  \\
\cline{3-11}
& & Replace Character & 1284 & 414 & 110 & 63 & 90.75 & 87.17 & 86.79 & 82.72  \\
\hline\hline
\multirow{9}{*}{DistilBERT}
& - & Attack Free & 1382 & 458 & 12 & 19 & 98.34 & 98.04 & 96.01 & 96.72 \\
\cline{2-11}
& \multirow{4}{*}{Word-Level}
& Out Of Vocab & 1343 & 348 & 51 & 129 & 90.37 & 89.22 & 72.95 & 79.45 \\
\cline{3-11}
& & Word Deletion & 1064 & 440 & 330 & 37 & 80.38 & 76.89 & 92.24 & 70.56 \\
\cline{3-11}
& & Synonym Replacement & 1388 & 417 & 6 & 60 & 96.47 & 97.21 & 87.42 & 92.66 \\
\cline{3-11}
& & Antonym Replacement & 1386 & 443 & 8 & 34 & 97.75 & 97.91 & 92.87 & 95.47 \\
\cline{2-11}
& \multirow{4}{*}{Character-Level} 
& Swap Letters & 1350 & 368 & 44 & 109 & 91.82 & 90.92 & 77.14 & 82.78 \\
\cline{3-11}
& & Delete Character & 1374 & 379 & 20 & 98 & 93.69 & 94.16 & 79.45 & 86.52 \\
\cline{3-11}
& & Insert Character & 1195 & 396 & 199 & 81 & 85.03 & 80.10 & 83.01 & 73.88 \\
\cline{3-11}
& & Replace Character & 1266 & 419 & 128 & 58 & 90.05 & 86.10 & 87.84 & 81.83 \\
\hline\hline
\end{tabular}}
\label{tab_awSpamAssassin}
\end{table*}

\begin{table*}[!t]
\centering
\caption{Attack Results for SpamAssassin Dataset with Replace One Score}
\resizebox{\textwidth}{!}{\begin{tabular}{c|c c c c c c c c c c}
\hline\hline
\textbf{Model} & \textbf{Attack Level} & \textbf{Attack} & \textbf{TP} & \textbf{TN} & \textbf{FP} & \textbf{FN} & \textbf{Accuracy} & \textbf{Precision} & \textbf{Recall} & \textbf{F1 Score}\rule[-2ex]{0pt}{6ex}\\
\hline\hline
\multirow{9}{*}{LSTM}
& - & Attack Free & 1393 & 468 & 1 & 9 & 99.46 & 99.57 & 98.11 & 98.94  \\
\cline{2-11}
& \multirow{4}{*}{Word-Level} 
& Out Of Vocab & 1387 & 316 & 7 & 61 & 91.02 & 93.72 & 66.25 & 79.00  \\
\cline{3-11}
& & Word Deletion & 1387 & 316 & 7 & 61 & 91.02 & 93.72 & 66.25 & 79.00  \\
\cline{3-11}
& & Synonym Replacement & 1376 & 451 & 18 & 26 & 97.65 & 97.15 & 94.55 & 95.35  \\
\cline{3-11}
& & Antonym Replacement & 1389 & 453 & 5 & 24 & 98.45 & 98.60 & 94.97 & 96.90  \\
\cline{2-11}
& \multirow{4}{*}{Character-Level} 
& Swap Letters & 1140 & 439 & 254 & 38 & 84.39 & 80.06 & 92.03 & 75.04  \\
\cline{3-11}
& & Delete Character & 1202 & 427 & 192 & 50 & 87.07 & 82.49 & 89.52 & 77.92  \\
\cline{3-11}
& & Insert Character & 1376 & 322 &   18 & 155 & 90.75 & 92.29 & 67.51 & 78.82  \\
\cline{3-11}
& & Replace Character & 1202 & 434 & 192  &  43 & 87.44 & 82.94 & 90.99 & 78.69  \\
\hline\hline
\multirow{9}{*}{Dense}
& - & Attack Free & 1391 & 464 & 3 & 13 & 99.14 & 99.21 & 97.27 & 98.30  \\
\cline{2-11}
& \multirow{4}{*}{Word-Level} 
& Out Of Vocab & 107 & 463 & 1287 & 14 & 30.46 & 57.44 & 97.06 & 41.58  \\
\cline{3-11}
& & Word Deletion & 106 & 463 & 1288 & 14 & 30.41 & 57.39 & 97.06 & 41.56  \\
\cline{3-11}
& & Synonym Replacement & 1377 & 466 & 17 & 11 & 98.50 & 97.84 & 97.69 & 97.08  \\
\cline{3-11}
& & Antonym Replacement & 1385 & 459 & 9 & 18 & 98.56 & 98.40 & 96.23 & 97.14  \\
\cline{2-11}
& \multirow{4}{*}{Character-Level} 
& Swap Letters & 607 & 436 & 787 & 41 & 55.75 & 64.66 & 91.40 & 51.29  \\
\cline{3-11}
& & Delete Character & 703 & 407 & 691 & 70 & 59.33 & 64.01 & 85.32 & 51.68  \\
\cline{3-11}
& & Insert Character & 113 & 450 & 1281 & 27 & 30.09 & 53.36 & 94.34 & 40.76  \\
\cline{3-11}
& & Replace Character & 595 & 440 & 799 & 37 & 55.32 & 64.83 & 92.24 & 51.28  \\
\hline\hline
\multirow{9}{*}{CNN}
& - & Attack Free & 1387 & 475 & 7 & 2 & 99.51 & 99.20 & 99.58 & 99.06  \\
\cline{2-11}
& \multirow{4}{*}{Word-Level} 
& Out Of Vocab & 860 & 383 &  534 & 94 & 66.44 & 65.96 & 80.29 & 54.95  \\
\cline{3-11}
& & Word Deletion & 861 & 383 & 533 & 94 & 66.49 & 65.98 & 80.29 & 54.99  \\
\cline{3-11}
& & Synonym Replacement & 1388 & 445 & 6 & 32 & 97.97 & 98.21 & 93.29 & 95.91  \\
\cline{3-11}
& & Antonym Replacement & 1387 & 457 & 7 & 20 & 98.56 & 98.53 & 95.81 & 97.13  \\
\cline{2-11}
& \multirow{4}{*}{Character-Level} 
& Swap Letters & 796 & 446 & 598 & 31 & 66.38 & 69.49 & 93.50 & 58.65  \\
\cline{3-11}
& & Delete Character & 931 & 412 & 463 & 65 & 71.78 & 70.28 & 86.37 & 60.95  \\
\cline{3-11}
& & Insert Character & 736 & 382 &  658 &  95 & 59.75 & 62.65 & 80.08 & 50.36  \\
\cline{3-11}
& & Replace Character & 813 & 442 & 581  &  35 & 67.08 & 69.54 & 92.66 & 58.93  \\
\hline\hline
\multirow{9}{*}{Attention}
& - & Attack Free & 1389 & 471 & 5 & 6 & 99.41 & 99.25 & 98.74 & 98.84  \\
\cline{2-11}
& \multirow{4}{*}{Word-Level}
& Out Of Vocab & 1340 & 382 & 54 & 95 & 92.04 & 90.50 & 80.08 & 83.68  \\
\cline{3-11}
& & Word Deletion & 1341 & 382 & 53 & 95 & 92.09 & 90.60 & 80.08 & 83.77  \\
\cline{3-11}
& & Synonym Replacement & 1389 & 457 & 5 & 20 & 98.66 & 98.75 & 95.81 & 97.34  \\
\cline{3-11}
& & Antonym Replacement & 1387 & 473 & 7 & 4 & 99.41 & 99.13 & 99.16 & 98.85  \\
\cline{2-11}
& \multirow{4}{*}{Character-Level} 
& Swap Letters & 701 & 471 & 693 & 6 & 62.64 & 69.81 & 98.74 & 57.40  \\
\cline{3-11}
& & Delete Character & 731 & 469 & 663 & 8 & 64.14 & 70.17 & 98.32 & 58.30  \\
\cline{3-11}
& & Insert Character & 1326 & 392 & 68 & 85 & 91.82 & 89.60 & 82.18 & 83.67  \\
\cline{3-11}
& & Replace Character & 774 & 468 & 620 & 9 & 66.38 & 70.93 & 98.11 & 59.81  \\
\hline\hline
\multirow{9}{*}{Transformer}
& - & Attack Free & 1389 & 460 & 5 & 17 & 98.82 & 98.86 & 96.44 & 97.66  \\
\cline{2-11}
& \multirow{4}{*}{Word-Level}
& Out Of Vocab & 1273 & 270 & 121 & 207 & 82.47 & 77.53 & 56.60 & 62.21  \\
\cline{3-11}
& & Word Deletion & 1274 & 270 & 120 & 207 & 82.52 & 77.63 & 56.60 & 62.28  \\
\cline{3-11}
& & Synonym Replacement & 1383 & 450 & 11 & 27 & 97.97 & 97.85 & 94.34 & 95.95  \\
\cline{3-11}
& & Antonym Replacement & 1388 & 458 &  6 & 19 & 98.66 & 98.68 & 96.02 & 97.34  \\
\cline{2-11}
& \multirow{4}{*}{Character-Level} 
& Swap Letters & 1194 & 387 & 200 & 90 & 84.50 & 79.46 & 81.13 & 72.74  \\
\cline{3-11}
& & Delete Character & 1219 & 361 & 175 &116 & 84.45 & 79.33 & 75.68 & 71.27  \\
\cline{3-11}
& & Insert Character & 1254 & 282 & 140 & 195 & 82.10 & 76.68 & 59.12 & 62.74  \\
\cline{3-11}
& & Replace Character & 1195 & 393 & 199 & 84 & 84.87 & 79.91 & 82.39 & 73.53  \\
\hline\hline
\multirow{9}{*}{DistilBERT}
& - & Attack Free & 1382 & 458 & 12 & 19 & 98.34 & 98.04 & 96.01 & 96.72 \\
\cline{2-11}
& \multirow{4}{*}{Word-Level}
& Out Of Vocab & 1357 & 266 & 37 & 211 & 86.74 & 87.16 & 55.76 & 68.2 \\
\cline{3-11}
& & Word Deletion & 1107 & 448 & 287 & 29 & 83.11 & 79.19 & 93.92 & 73.92 \\
\cline{3-11}
& & Synonym Replacement & 1384 & 432 & 10 & 45 & 97.06 & 97.29 & 90.56 & 94.01 \\
\cline{3-11}
& & Antonym Replacement & 1384 & 450 & 10 & 27 & 98.01 & 97.95 & 94.33 & 96.05 \\
\cline{2-11}
& \multirow{4}{*}{Character-Level} 
& Swap Letters & 1352 & 354 & 42 & 123 & 91.18 & 90.52 & 74.21 & 81.09 \\
\cline{3-11}
& & Delete Character & 1368 & 377 & 26 & 100 & 93.26 & 93.36 & 79.03 & 85.68 \\
\cline{3-11}
& & Insert Character & 1028 & 422 & 366 & 55 & 77.49 & 74.23 & 88.46 & 66.71 \\
\cline{3-11}
& & Replace Character & 1184 & 421 & 210 & 56 & 85.78 & 81.10 & 88.25 & 75.99 \\
\hline\hline
\end{tabular}}
\label{tab_r1SpamAssassin}
\end{table*}

\begin{table*}[!t]
\centering
\caption{Attack Results for TREC 2007 Dataset with Spam Weights}
\resizebox{\textwidth}{!}{\begin{tabular}{c|c c c c c c c c c c}
\hline\hline
\textbf{Model} & \textbf{Attack Level} & \textbf{Attack} & \textbf{TP} & \textbf{TN} & \textbf{FP} & \textbf{FN} & \textbf{Accuracy} & \textbf{Precision} & \textbf{Recall} & \textbf{F1 Score}\rule[-2ex]{0pt}{6ex}\\
\hline\hline
\multirow{9}{*}{LSTM}
& - & Attack Free & 5035 & 9958 & 38 & 53 & 99.40 & 99.29 & 99.47 & 99.55  \\
\cline{2-11}
& \multirow{4}{*}{Word-Level} 
& Out Of Vocab & 1772 & 10011 & 3301 & 0 & 78.12 & 87.60 & 100.00 & 85.85  \\
\cline{3-11}
& & Word Deletion & 1887 & 10010 & 3186 &  1 & 78.87 & 87.90 & 99.99 & 86.27  \\
\cline{3-11}
& & Synonym Replacement & 5000 & 9950 & 73 & 61 & 99.11 & 99.03 & 99.39 & 99.33  \\
\cline{3-11}
& & Antonym Replacement & 5062 & 9925 & 11 & 86 & 99.36 & 99.11 & 99.14 & 99.51 
  \\
\cline{2-11}
& \multirow{4}{*}{Character-Level} 
& Swap Letters & 3649 & 10003 & 1424 &  8 & 90.51 & 93.66 & 99.92 & 93.32  \\
\cline{3-11}
& & Delete Character & 3482 & 10006 & 1591 & 5 & 89.42 & 93.07 & 99.95 & 92.61  \\
\cline{3-11}
& & Insert Character & 1849 & 10010 & 3224 & 1 & 78.62 & 87.79 & 99.99 & 86.13 \\
\cline{3-11}
& & Replace Character & 3649 & 10006 & 1424 & 5 & 90.53 & 93.70 & 99.95 & 93.34  \\
\hline\hline
\multirow{9}{*}{Dense}
& - & Attack Free & 5036 & 9973 & 37 & 38 & 99.50 & 99.44 & 99.62 & 99.63  \\
\cline{2-11}
& \multirow{4}{*}{Word-Level} 
& Out Of Vocab &  4366 & 9997 & 707 & 14 & 95.22 & 96.54 & 99.86 & 96.52  \\
\cline{3-11}
& & Word Deletion & 4099 & 9998 & 974 & 13 & 93.46 & 95.40 & 99.87 & 95.30  \\
\cline{3-11}
& & Synonym Replacement & 4988 & 9984 & 85 & 27 & 99.26 & 99.31 & 99.73 & 99.44  \\
\cline{3-11}
& & Antonym Replacement & 5034 & 9961 & 39 & 50 & 99.41 & 99.31 & 99.50 & 99.56  \\
\cline{2-11}
& \multirow{4}{*}{Character-Level} 
& Swap Letters & 4540 &  9994 & 533 & 17 & 96.35 & 97.28 & 99.83 & 97.32  \\
\cline{3-11}
& & Delete Character & 4505 & 9995 & 568 & 16 & 96.13 & 97.13 & 99.84 & 97.16  \\
\cline{3-11}
& & Insert Character & 4103 & 10000 & 970 & 11 & 93.50 & 95.45 & 99.89 & 95.32  \\
\cline{3-11}
& & Replace Character & 4540 & 9992 & 533 & 19 & 96.34 & 97.26 & 99.81 & 97.31  \\
\hline\hline
\multirow{9}{*}{CNN}
& - & Attack Free & 5034 & 9987 & 39 & 24 & 99.58 & 99.57 & 99.76 & 99.69  \\
\cline{2-11}
& \multirow{4}{*}{Word-Level} 
& Out Of Vocab & 3995 &  9993 & 1078 & 18 & 92.73 & 94.91 & 99.82 & 94.80  \\
\cline{3-11}
& & Word Deletion & 3563 & 9992 & 1510 & 19 & 89.86 & 93.17 & 99.81 & 92.89  \\
\cline{3-11}
& & Synonym Replacement & 5037 & 9980 & 36 & 31 & 99.56 & 99.51 & 99.69 & 99.67 \\
\cline{3-11}
& & Antonym Replacement & 5043 & 9969 & 30 & 42 & 99.52 & 99.44 & 99.58 & 99.64  \\
\cline{2-11}
& \multirow{4}{*}{Character-Level} 
& Swap Letters & 4386 & 9990 & 687 & 21 & 95.31 & 96.54 & 99.79 & 96.58  \\
\cline{3-11}
& & Delete Character & 4342 & 9990 & 731 &  21 & 95.01 & 96.35 & 99.79 & 96.37  \\
\cline{3-11}
& & Insert Character & 3608 & 9992 & 1465 & 19 & 90.16 & 93.34 & 99.81 & 93.09  \\
\cline{3-11}
& & Replace Character &  4385 & 9986 & 688 & 25 & 95.27 & 96.49 & 99.75 & 96.55  \\
\hline\hline
\multirow{9}{*}{Attention}
& - & Attack Free & 5037 & 9997 & 36 & 14 & 99.67 & 99.68 & 99.86 & 99.75  \\
\cline{2-11}
& \multirow{4}{*}{Word-Level}
& Out Of Vocab & 3589 & 10006 & 1484 & 5 & 90.13 & 93.47 & 99.95 & 93.07  \\
\cline{3-11}
& & Word Deletion & 3488 & 10007 & 1585 & 4 & 89.47 & 93.11 & 99.96 & 92.64  \\
\cline{3-11}
& & Synonym Replacement & 5015 & 9976 & 58 & 35 & 99.38 & 99.36 & 99.65 & 99.54  \\
 \cline{3-11}
& & Antonym Replacement & 5035 & 9981 & 38 & 30 & 99.55 & 99.51 & 99.70 & 99.66  \\
\cline{2-11}
& \multirow{4}{*}{Character-Level} 
& Swap Letters & 4340 & 10003 & 733 & 8 & 95.09 & 96.49 & 99.92 & 96.43  \\
\cline{3-11}
& & Delete Character & 4218 & 10003 & 855 & 8 & 94.28 & 95.97 & 99.92 & 95.86  \\
\cline{3-11}
& & Insert Character & 3444 & 10007 & 1629 & 4 & 89.17 & 92.94 & 99.96 & 92.46  \\
\cline{3-11}
& & Replace Character & 4285 & 10002 & 788 & 9 & 94.72 & 96.24 & 99.91 & 96.17  \\
\hline\hline
\multirow{9}{*}{Transformer}
& - & Attack Free &  4895 & 9963 & 178 & 48 & 98.50 & 98.64 & 99.52 & 98.88  \\
\cline{2-11}
& \multirow{4}{*}{Word-Level}
& Out Of Vocab & 4236 & 9983 & 837 & 28 & 94.27 & 95.80 & 99.72 & 95.85  \\
\cline{3-11}
& & Word Deletion & 3195 & 9994 & 1878 & 17 & 87.44 & 91.83 & 99.83 & 91.34  \\
\cline{3-11}
& & Synonym Replacement & 4719 & 9978 & 354 & 33 & 97.43 & 97.94 & 99.67 & 98.10  \\
\cline{3-11}
& & Antonym Replacement & 4887 & 9968 & 186 & 43 & 98.48 & 98.65 & 99.57 & 98.86  \\
\cline{2-11}
& \multirow{4}{*}{Character-Level} 
& Swap Letters & 3779 & 9988 & 1294 & 23 & 91.27 & 93.96 & 99.77 & 93.81  \\
\cline{3-11}
& & Delete Character & 3733 & 9987 & 1340 & 24 & 90.96 & 93.77 & 99.76 & 93.61  \\
\cline{3-11}
& & Insert Character & 3248 & 9995 & 1825 & 16 & 87.80 & 92.03 & 99.84 & 91.57  \\
\cline{3-11}
& & Replace Character & 3750 & 9988 & 1323 & 23 & 91.08 & 93.85 & 99.77 & 93.69  \\
\hline\hline
\multirow{9}{*}{DistilBERT}
& - & Attack Free & 5061 & 10001 & 12 & 10 & 99.85 & 99.84 & 99.90 & 99.89 \\
\cline{2-11}
& \multirow{4}{*}{Word-Level}
& Out Of Vocab & 4868 & 10001 & 205 & 10 & 98.57 & 98.89 & 99.90 & 98.94 \\
\cline{3-11}
& & Word Deletion & 5060 & 10000 & 13 & 11 & 99.84 & 99.83 & 99.89 & 99.88 \\
\cline{3-11}
& & Synonym Replacement & 5013 & 10002 & 60 & 9 & 99.54 & 99.61 & 99.91 & 99.66 \\
\cline{3-11}
& & Antonym Replacement & 5058 & 10000 & 15 & 11 & 99.83 & 99.82 & 99.89 & 99.87 \\
\cline{2-11}
& \multirow{4}{*}{Character-Level} 
& Swap Letters & 4682 & 10004 & 391 & 7 & 97.36 & 98.04 & 99.93 & 98.05 \\
\cline{3-11}
& & Delete Character & 4772 & 10003 & 301 & 8 & 97.95 & 98.46 & 99.92 & 98.48 \\
\cline{3-11}
& & Insert Character & 4512 & 10005 & 561 & 6 & 96.24 & 97.28 & 99.94 & 97.24 \\
\cline{3-11}
& & Replace Character & 4746 & 10002 & 327 & 9 & 97.77 & 98.32 & 99.91 & 98.35 \\
\hline\hline
\end{tabular}}
\label{tab_swTrec}
\end{table*}

\begin{table*}[!t]
\centering
\caption{Attack Results for TREC 2007 Dataset with Attention Weights}
\resizebox{\textwidth}{!}{\begin{tabular}{c|c c c c c c c c c c}
\hline\hline
\textbf{Model} & \textbf{Attack Level} & \textbf{Attack} & \textbf{TP} & \textbf{TN} & \textbf{FP} & \textbf{FN} & \textbf{Accuracy} & \textbf{Precision} & \textbf{Recall} & \textbf{F1 Score}\rule[-2ex]{0pt}{6ex}\\
\hline\hline
\multirow{9}{*}{LSTM}
& - & Attack Free & 5035 & 9958 & 38 & 53 & 99.40 & 99.29 & 99.47 & 99.55  \\
\cline{2-11}
& \multirow{4}{*}{Word-Level} 
& Out Of Vocab & 4944 & 9978 & 129 & 33 & 98.93 &  99.03 & 99.67 &  99.19  \\
\cline{3-11}
& & Word Deletion & 4965 & 9974 & 108 & 37 & 99.04 &  99.09 & 99.63 &  99.28  \\
\cline{3-11}
& & Synonym Replacement & 5003 & 9951 & 70 & 60 & 99.14 & 99.06 & 99.40 & 99.35  \\
\cline{3-11}
& & Antonym Replacement & 5009 & 9973 & 64 & 38 & 99.32 & 99.30 & 99.62 & 99.49  \\
\cline{2-11}
& \multirow{4}{*}{Character-Level} 
& Swap Letters & 4972 & 9964 & 101 & 47 & 99.02 &  99.03 & 99.53 & 99.26  \\
\cline{3-11}
& & Delete Character & 4956 & 9973 & 117 & 38 & 98.97 & 99.04 & 99.62 & 99.23  \\
\cline{3-11}
& & Insert Character & 4958 & 9971 & 115 & 40 & 98.97 & 99.03 & 99.60 & 99.23  \\
\cline{3-11}
& & Replace Character & 4964 & 9961 & 109 & 50 & 98.95 & 98.96 & 99.50 & 99.21  \\
\hline\hline
\multirow{9}{*}{Dense}
& - & Attack Free & 5036 & 9973 & 37 & 38 & 99.50 & 99.44 & 99.62 & 99.63  \\
\cline{2-11}
& \multirow{4}{*}{Word-Level} 
& Out Of Vocab & 5024 & 9983 &  49 & 28 & 99.49 &  99.48 & 99.72 &  99.62  \\
\cline{3-11}
& & Word Deletion &  4976 & 9989 &  97 & 22 & 99.21 &  99.30 & 99.78 &  99.41  \\
\cline{3-11}
& & Synonym Replacement & 4986 & 9987 & 87 & 24 & 99.26 & 99.33 & 99.76 & 99.45  \\
\cline{3-11}
& & Antonym Replacement & 5035 & 9966 & 38 & 45 & 99.45 & 99.37 & 99.55 & 99.59  \\
\cline{2-11}
& \multirow{4}{*}{Character-Level} 
& Swap Letters & 4985 & 9992 & 88 & 19 & 99.29 & 99.37 & 99.81 & 99.47  \\
\cline{3-11}
& & Delete Character & 4998 & 9988 & 75 & 23 & 99.35 & 99.40 & 99.77 & 99.51  \\
\cline{3-11}
& & Insert Character & 4983 & 9986 & 90 & 25 & 99.24 & 99.30 & 99.75 & 99.43  \\
\cline{3-11}
& & Replace Character & 4979 & 9993 & 94 & 18 & 99.26 & 99.35 & 99.82 & 99.44  \\
\hline\hline
\multirow{9}{*}{CNN}
& - & Attack Free & 5034 & 9987 & 39 & 24 & 99.58 & 99.57 & 99.76 & 99.69  \\
\cline{2-11}
& \multirow{4}{*}{Word-Level} 
& Out Of Vocab & 4995 & 9964 &  78 & 47 & 99.17 &  99.15 & 99.53 &  99.38  \\
\cline{3-11}
& & Word Deletion &  4953 & 9975 & 120 & 36 & 98.97 &  99.04 & 99.64 &  99.22  \\
\cline{3-11}
& & Synonym Replacement & 5061 & 9929 & 12 & 82 & 99.38 & 99.14 & 99.18 & 99.53  \\
\cline{3-11}
& & Antonym Replacement & 5039 & 9978 & 34 & 33 & 99.56 & 99.50 & 99.67 & 99.67    \\
\cline{2-11}
& \multirow{4}{*}{Character-Level} 
& Swap Letters & 4989 & 9981 & 84 & 30 & 99.24 & 99.28 & 99.70 & 99.43  \\
\cline{3-11}
& & Delete Character & 4976 & 9988 & 97 & 23 & 99.20 & 99.29 & 99.77 & 99.40  \\
\cline{3-11}
& & Insert Character & 4957 & 9975 & 116 & 36 & 98.99 & 99.06 & 99.64 & 99.24  \\
\cline{3-11}
& & Replace Character & 4989 & 9981 &  84 & 30 & 99.24 & 99.28 & 99.70 & 99.43  \\
\hline\hline
\multirow{9}{*}{Attention}
& - & Attack Free & 5037 & 9997 & 36 & 14 & 99.67 & 99.68 & 99.86 & 99.75  \\
\cline{2-11}
& \multirow{4}{*}{Word-Level}
& Out Of Vocab & 4998 & 9992 & 75 & 19 & 99.38 & 99.44 & 99.81 & 99.53  \\
\cline{3-11}
& & Word Deletion & 5002 & 9994 & 71 & 17 & 99.42 & 99.48 & 99.83 & 99.56  \\
\cline{3-11}
& & Synonym Replacement & 5015 & 9977 & 58 & 34 & 99.39 & 99.37 & 99.66 & 99.54  \\
 \cline{3-11}
& & Antonym Replacement & 5033 & 9995 & 40 & 16 & 99.63 & 99.64 & 99.84 & 99.72  \\
\cline{2-11}
& \multirow{4}{*}{Character-Level} 
& Swap Letters & 5010 & 9998 &  63 & 13 & 99.50 &  99.56 & 99.87 &  99.62  \\
\cline{3-11}
& & Delete Character & 4988 & 9998 &  85 & 13 & 99.35 &  99.45 & 99.87 &  99.51  \\
\cline{3-11}
& & Insert Character & 4997 & 9993 &  76 & 18 & 99.38 &  99.44 & 99.82 &  99.53  \\
\cline{3-11}
& & Replace Character & 5009 & 9998 &  64 & 13 & 99.49 &  99.55 & 99.87 &  99.62  \\
\hline\hline
\multirow{9}{*}{Transformer}
& - & Attack Free &  4895 & 9963 & 178 & 48 & 98.50 & 98.64 & 99.52 & 98.88  \\
\cline{2-11}
& \multirow{4}{*}{Word-Level}
& Out Of Vocab & 4977 & 9204 & 96 & 807 & 94.01 & 92.51 & 91.94 & 95.32  \\
\cline{3-11}
& & Word Deletion & 4678 & 9970 & 395 & 41 & 97.11 & 97.66 & 99.59 & 97.86  \\
\cline{3-11}
& & Synonym Replacement &  4856 & 9978 & 217 & 33 & 98.34 & 98.60 & 99.67 & 98.76  \\
\cline{3-11}
& & Antonym Replacement & 4892 & 9963 & 181 & 48 & 98.48 & 98.62 & 99.52 & 98.86  \\
\cline{2-11}
& \multirow{4}{*}{Character-Level} 
& Swap Letters & 4749 & 9970 & 324 & 41 & 97.58 & 98.00 & 99.59 & 98.20  \\
\cline{3-11}
& & Delete Character & 4743 & 9961 & 330 & 50 & 97.48 & 97.88 & 99.50 & 98.13  \\
\cline{3-11}
& & Insert Character & 4685 & 9972 & 388 & 39 & 97.17 & 97.71 & 99.61 & 97.90  \\
\cline{3-11}
& & Replace Character & 4717 & 9970 & 356 & 41 & 97.37 & 97.85 & 99.59 & 98.05  \\
\hline\hline
\multirow{9}{*}{DistilBERT}
& - & Attack Free & 5061 & 10001 & 12 & 10 & 99.85 & 99.84 & 99.90 & 99.89 \\
\cline{2-11}
& \multirow{4}{*}{Word-Level}
& Out Of Vocab & 5060 & 9996 & 13 & 15 & 99.81 & 99.79 & 99.85 & 99.86 \\
\cline{3-11}
& & Word Deletion & 5057 & 9996 & 16 & 15 & 99.79 & 99.77 & 99.85 & 99.85 \\
\cline{3-11}
& & Synonym Replacement & 5059 & 9997 & 14 & 14 & 99.81 & 99.79 & 99.86 & 99.86 \\
\cline{3-11}
& & Antonym Replacement & 5062 & 9999 & 11 & 12 & 99.85 & 99.83 & 99.88 & 99.89 \\
\cline{2-11}
& \multirow{4}{*}{Character-Level} 
& Swap Letters & 5046 & 9996 & 27 & 15 & 99.72 & 99.72 & 99.85 & 99.79 \\
\cline{3-11}
& & Delete Character & 5049 & 9997 & 24 & 14 & 99.75 & 99.74 & 99.86 & 99.81 \\
\cline{3-11}
& & Insert Character & 5032 & 10000 & 41 & 11 & 99.66 & 99.69 & 99.89 & 99.74 \\
\cline{3-11}
& & Replace Character & 5045 & 9997 & 28 & 14 & 99.72 & 99.72 & 99.86 & 99.79 \\
\hline\hline
\end{tabular}}
\label{tab_awTrec}
\end{table*}

\begin{table*}[!t]
\centering
\caption{Attack Results for TREC 2007 Dataset with Replace One Score}
\resizebox{\textwidth}{!}{\begin{tabular}{c|c c c c c c c c c c}
\hline\hline
\textbf{Model} & \textbf{Attack Level} & \textbf{Attack} & \textbf{TP} & \textbf{TN} & \textbf{FP} & \textbf{FN} & \textbf{Accuracy} & \textbf{Precision} & \textbf{Recall} & \textbf{F1 Score}\rule[-2ex]{0pt}{6ex}\\
\hline\hline
\multirow{9}{*}{LSTM}
& - & Attack Free & 5035 & 9958 & 38 & 53 & 99.40 & 99.29 & 99.47 & 99.55  \\
\cline{2-11}
& \multirow{4}{*}{Word-Level} 
& Out Of Vocab & 4736 & 9939 & 337 & 72 & 97.29 & 97.61 & 99.28 & 97.98  \\
\cline{3-11}
& & Word Deletion & 4915 & 9936 & 158 & 75 & 98.46 & 98.47 & 99.25 & 98.84  \\
\cline{3-11}
& & Synonym Replacement & 5002 & 9955 & 71 & 56 & 99.16 & 99.09 & 99.44 & 99.37  \\
\cline{3-11}
& & Antonym Replacement & 5013 & 9976 & 60 & 35 & 99.37 & 99.35 & 99.65 & 99.53  \\
\cline{2-11}
& \multirow{4}{*}{Character-Level} 
& Swap Letters & 4980 & 9924 &  93 &  87 & 98.81 & 98.68 & 99.13 & 99.10  \\
\cline{3-11}
& & Delete Character & 4931 & 9880 & 142 & 131 & 98.19 & 98.00 & 98.69 & 98.64  \\
\cline{3-11}
& & Insert Character & 4897 & 9843 & 176 & 168 & 97.72 & 97.46 & 98.32 & 98.28  \\
\cline{3-11}
& & Replace Character & 4982 & 9918 & 91 & 93 & 98.78 & 98.63 & 99.07 & 99.08  \\
\hline\hline
\multirow{9}{*}{Dense}
& - & Attack Free & 5036 & 9973 & 37 & 38 & 99.50 & 99.44 & 99.62 & 99.63  \\
\cline{2-11}
& \multirow{4}{*}{Word-Level} 
& Out Of Vocab & 4869 & 9827 & 204 & 184 & 97.43 & 97.16 & 98.16 & 98.06  \\
\cline{3-11}
& & Word Deletion & 4652 & 9922 & 421 & 89 & 96.62 & 97.03 & 99.11 & 97.49  \\
\cline{3-11}
& & Synonym Replacement & 4986 & 9990 & 87 & 21 & 99.28 & 99.36 & 99.79 & 99.46  \\
\cline{3-11}
& & Antonym Replacement & 5057 & 9924 & 16 & 87 & 99.32 & 99.07 & 99.13 & 99.48  \\
\cline{2-11}
& \multirow{4}{*}{Character-Level} 
& Swap Letters & 4953 & 9936 & 120 & 75 & 98.71 & 98.66 & 99.25 & 99.03  \\
\cline{3-11}
& & Delete Character & 4927 & 9940 & 146 & 71 & 98.56 & 98.57 & 99.29 & 98.92  \\
\cline{3-11}
& & Insert Character & 4621 & 9882 & 452 & 129 & 96.15 & 96.46 & 98.71 & 97.14  \\
\cline{3-11}
& & Replace Character & 4949 & 9945 & 124 & 66 & 98.74 & 98.73 & 99.34 & 99.05  \\
\hline\hline
\multirow{9}{*}{CNN}
& - & Attack Free & 5034 & 9987 & 39 & 24 & 99.58 & 99.57 & 99.76 & 99.69  \\
\cline{2-11}
& \multirow{4}{*}{Word-Level} 
& Out Of Vocab & 4894 & 9266 & 179 & 745 & 93.87 & 92.45 & 92.56 & 95.25  \\
\cline{3-11}
& & Word Deletion & 4444 & 9934 & 629 & 77 & 95.32 & 96.17 & 99.23 & 96.57 \\
\cline{3-11}
& & Synonym Replacement &  5061 & 9925 & 12 & 86 & 99.35 & 99.10 & 99.14 & 99.51  \\
\cline{3-11}
& & Antonym Replacement & 5036 & 9978 & 37 & 33 & 99.54 & 99.49 & 99.67 & 99.65   \\
\cline{2-11}
& \multirow{4}{*}{Character-Level} 
& Swap Letters & 4932 & 9841 & 141 & 170 & 97.94 & 97.63 & 98.30 & 98.44  \\
\cline{3-11}
& & Delete Character & 4899 & 9819 & 174 & 192 & 97.57 & 97.24 & 98.08 & 98.17  \\
\cline{3-11}
& & Insert Character & 4483 & 9837 & 590 & 174 & 94.94 & 95.30 & 98.26 & 96.26  \\
\cline{3-11}
& & Replace Character &4931 & 9832 & 142 & 179 & 97.87 & 97.54 & 98.21 & 98.39  \\
\hline\hline
\multirow{9}{*}{Attention}
& - & Attack Free & 5037 & 9997 & 36 & 14 & 99.67 & 99.68 & 99.86 & 99.75  \\
\cline{2-11}
& \multirow{4}{*}{Word-Level}
& Out Of Vocab & 4953 & 9944 & 120 &  67 & 98.76 & 98.74 & 99.33 & 99.07  \\
\cline{3-11}
& & Word Deletion & 4964 & 9967 & 109 &  44 & 98.99 & 99.02 & 99.56 & 99.24  \\
\cline{3-11}
& & Synonym Replacement & 5017 & 9975 & 56 & 36 & 99.39 & 99.36 & 99.64 & 99.54  \\
 \cline{3-11}
& & Antonym Replacement & 5041 & 9988 & 32 & 23 & 99.64 & 99.61 & 99.77 & 99.73  \\
\cline{2-11}
& \multirow{4}{*}{Character-Level} 
& Swap Letters & 4991 & 9979 &  82 & 32 & 99.24 & 99.27 & 99.68 & 99.43  \\
\cline{3-11}
& & Delete Character & 4957 & 9942 & 116 & 69 & 98.77 & 98.74 & 99.31 & 99.08  \\
\cline{3-11}
& & Insert Character & 4952 & 9837 & 121 & 174 & 98.04 & 97.70 & 98.26 & 98.52  \\
\cline{3-11}
& & Replace Character & 4992 & 9971 &  81 &  40 & 99.20 & 99.20 & 99.60 & 99.40  \\
\hline\hline
\multirow{9}{*}{Transformer}
& - & Attack Free &  4895 & 9963 & 178 & 48 & 98.50 & 98.64 & 99.52 & 98.88  \\
\cline{2-11}
& \multirow{4}{*}{Word-Level}
& Out Of Vocab & 4883 & 6911 & 190 &3100 & 78.19 & 79.25 & 69.03 & 80.77  \\
\cline{3-11}
& & Word Deletion & 4084 & 9615 & 989 & 396 & 90.82 & 90.92 & 96.04 & 93.28  \\
\cline{3-11}
& & Synonym Replacement & 4842 & 9920 & 231 & 91 & 97.87 & 97.94 & 99.09 & 98.40  \\
\cline{3-11}
& & Antonym Replacement & 4856 & 9978 & 217 & 33 & 98.34 & 98.60 & 99.67 & 98.76  \\
\cline{2-11}
& \multirow{4}{*}{Character-Level} 
& Swap Letters & 4744 & 9606 & 329 & 405 & 95.13 & 94.41 & 95.95 & 96.32  \\
\cline{3-11}
& & Delete Character & 4689 & 9365 & 384 & 646 & 93.17 & 91.98 & 93.55 & 94.79  \\
\cline{3-11}
& & Insert Character & 4166 & 9558 & 907 & 453 & 90.98 & 90.76 & 95.47 & 93.36  \\
\cline{3-11}
& & Replace Character & 4695 & 9576 & 378 & 435 & 94.61 & 93.86 & 95.65 & 95.93  \\
\hline\hline
\multirow{9}{*}{DistilBERT}
& - & Attack Free & 5061 & 10001 & 12 & 10 & 99.85 & 99.84 & 99.90 & 99.89 \\
\cline{2-11}
& \multirow{4}{*}{Word-Level}
& Out Of Vocab & 5029 & 9981 & 44 & 30 & 99.51 & 99.48 & 99.70 & 99.63 \\
\cline{3-11}
& & Word Deletion & 4969 & 10003 & 104 & 8 & 99.26 & 99.41 & 99.92 & 99.44 \\
\cline{3-11}
& & Synonym Replacement & 5027 & 9993 & 46 & 18 & 99.58 & 99.59 & 99.82 & 99.68 \\
\cline{3-11}
& & Antonym Replacement & 5061 & 10001 & 12 & 10 & 99.85 & 99.84 & 99.90 & 99.89 \\
\cline{2-11}
& \multirow{4}{*}{Character-Level} 
& Swap Letters & 4780 & 10004 & 293 & 7 & 98.01 & 98.50 & 99.93 & 98.52 \\
\cline{3-11}
& & Delete Character & 4896 & 9998 & 177 & 13 & 98.74 & 99.00 & 99.87 & 99.06 \\
\cline{3-11}
& & Insert Character & 4699 & 10001 & 374 & 10 & 97.45 & 98.09 & 99.90 & 98.12 \\
\cline{3-11}
& & Replace Character & 4842 & 10000 & 231 & 11 & 98.40 & 98.76 & 99.89 & 98.80 \\
\hline\hline
\end{tabular}}
\label{tab_r1Trec}
\end{table*}

\begin{table*}[!t]
\centering
\caption{Attack Results of SpamAssassin Dataset with Spam Weights for Sentence-Level}
\resizebox{\textwidth}{!}{\begin{tabular}{c|c c c c c c c c c c}
\hline\hline
\textbf{Model} & \textbf{Attack Level} & \textbf{Attack} & \textbf{TP} & \textbf{TN} & \textbf{FP} & \textbf{FN} & \textbf{Accuracy} & \textbf{Precision} & \textbf{Recall} & \textbf{F1 Score}\rule[-2ex]{0pt}{6ex}\\
\hline\hline
\multirow{4}{*}{LSTM}
& - & Attack Free & 1393 & 468 & 1 & 9 & 99.46 & 99.57 & 98.11 & 98.94  \\
\cline{2-11}
& \multirow{3}{*}{Sentence-Level} 
& Add Ham Sentence & 1384 & 460 & 10 & 17 & 98.56 & 98.33 & 96.44 & 97.15  \\
\cline{3-11}
&  & Add Spam Sentence & 1381 & 461 & 13 & 16 & 98.45 & 98.06 & 96.65 & 96.95  \\
\cline{3-11}
&  & Add Ham-Spam Sentence & 1384 & 458 & 10 & 19 & 98.45 & 98.25 & 96.02 & 96.93  \\
\hline\hline
\multirow{4}{*}{Dense}
& - & Attack Free & 1391 & 464 & 3 & 13 & 99.14 & 99.21 & 97.27 & 98.30  \\
\cline{2-11}
& \multirow{3}{*}{Sentence-Level} 
& Add Ham Sentence & 1277 & 475 & 117 & 2 & 93.64 & 90.04 & 99.58 & 88.87  \\
\cline{3-11}
&  & Add Spam Sentence & 1282 & 433 & 112 & 44 & 91.66 & 88.07 & 90.78 & 84.74  \\
\cline{3-11}
&  & Add Ham-Spam Sentence & 1277 & 414 & 117 &  63 & 90.38 & 86.63 & 86.79 & 82.14  \\
\hline\hline
\multirow{4}{*}{CNN}
& - & Attack Free & 1387 & 475 & 7 & 2 & 99.51 & 99.20 & 99.58 & 99.06  \\
\cline{2-11}
& \multirow{3}{*}{Sentence-Level} 
& Add Ham Sentence & 1355 & 465 & 39 & 12 & 97.27 & 95.69 & 97.48 & 94.80  \\
\cline{3-11}
&  & Add Spam Sentence & 1347 & 440 & 47 & 37 & 95.51 & 93.84 & 92.24 & 91.29  \\
\cline{3-11}
&  & Add Ham-Spam Sentence & 1355 & 226 & 39 & 251 & 84.50 & 84.83 & 47.38 & 60.92  \\
\hline\hline
\multirow{4}{*}{Attention}
& - & Attack Free & 1389 & 471 & 5 & 6 & 99.41 & 99.25 & 98.74 & 98.84  \\
\cline{2-11}
& \multirow{3}{*}{Sentence-Level} 
& Add Ham Sentence & 1392 & 468 &   2 &  9 & 99.41 & 99.47 & 98.11 & 98.84  \\
\cline{3-11}
&  & Add Spam Sentence & 1392 & 467 & 2 & 10 & 99.36 & 99.43 & 97.90 & 98.73  \\
\cline{3-11}
&  & Add Ham-Spam Sentence & 1392 & 467 & 2 &  10 & 99.36 & 99.43 & 97.90 & 98.73  \\
\hline\hline
\multirow{4}{*}{Transformer}
& - & Attack Free & 1389 & 460 & 5 & 17 & 98.82 & 98.86 & 96.44 & 97.66  \\
\cline{2-11}
& \multirow{3}{*}{Sentence-Level} 
& Add Ham Sentence & 1389 & 461 & 5 & 16 & 98.88 & 98.89 & 96.65 & 97.77  \\
\cline{3-11}
&  & Add Spam Sentence & 1386 & 460 & 8 & 17 & 98.66 & 98.54 & 96.44 & 97.35  \\
\cline{3-11}
&  & Add Ham-Spam Sentence & 1389 & 460 & 5 & 17 & 98.82 & 98.86 & 96.44 & 97.66  \\
\hline\hline
\multirow4{*}{DistilBERT}
& - & Attack Free & 1382 & 458 & 12 & 19 & 98.34 & 98.04 & 96.01 & 96.72 \\
\cline{2-11}
& \multirow{3}{*}{Sentence-Level} 
& Add Ham Sentence & 1388 & 437 & 6 & 40 & 97.54 & 97.92 & 91.61 & 94.99 \\
\cline{3-11}
&  & Add Spam Sentence & 1385 & 447 & 9 & 30 & 97.91 & 97.95 & 93.71 & 95.81 \\
\cline{3-11}
&  & Add Ham-Spam Sentence & 1382 & 452 & 12 & 25 & 98.02 & 97.81 & 94.75 & 96.06 \\
\hline\hline
\end{tabular}}
\label{tab_swSentenceSpamAssassin}
\end{table*}

\begin{table*}[!t]
\centering
\caption{Attack Results of TREC 2007 Dataset with Spam Weights for Sentence-Level}
\resizebox{\textwidth}{!}{\begin{tabular}{c|c c c c c c c c c c}
\hline\hline
\textbf{Model} & \textbf{Attack Level} & \textbf{Attack} & \textbf{TP} & \textbf{TN} & \textbf{FP} & \textbf{FN} & \textbf{Accuracy} & \textbf{Precision} & \textbf{Recall} & \textbf{F1 Score}\rule[-2ex]{0pt}{6ex}\\
\hline\hline
\multirow{4}{*}{LSTM}
& - & Attack Free & 5035 & 9958 & 38 & 53 & 99.40 & 99.29 & 99.47 & 99.55  \\
\cline{2-11}
& \multirow{3}{*}{Sentence-Level} 
& Add Ham Sentence & 4953 &  9983 & 120 & 28 & 99.02 & 99.13 & 99.72 & 99.26  \\
\cline{3-11}
&  & Add Spam Sentence & 5039 & 9442 & 34 &  569 & 96.00 & 94.75 & 94.32 & 96.91  \\
\cline{3-11}
&  & Add Ham-Spam Sentence & 4953 & 9442 & 120 & 569 & 95.43 & 94.22 & 94.32 & 96.48  \\
\hline\hline
\multirow{4}{*}{Dense}
& - & Attack Free & 5036 & 9973 & 37 & 38 & 99.50 & 99.44 & 99.62 & 99.63  \\
\cline{2-11}
& \multirow{3}{*}{Sentence-Level} 
& Add Ham Sentence & 4291 & 10002 & 782 & 9 & 94.76 & 96.27 & 99.91 & 96.20  \\
\cline{3-11}
&  & Add Spam Sentence & 5062 & 3430 & 11 & 6581 & 56.30 & 71.58 & 34.26 & 51.00  \\
\cline{3-11}
&  & Add Ham-Spam Sentence &  4291 & 3430 & 782 & 6581 & 51.19 & 60.45 & 34.26 & 48.23  \\
\hline\hline
\multirow{4}{*}{CNN}
& - & Attack Free & 5034 & 9987 & 39 & 24 & 99.58 & 99.57 & 99.76 & 99.69  \\
\cline{2-11}
& \multirow{3}{*}{Sentence-Level} 
& Add Ham Sentence & 5044 & 9926 & 29 & 85 & 99.24 & 99.03 & 99.15 & 99.43  \\
\cline{3-11}
&  & Add Spam Sentence & 5067 & 7854 & 6 & 2157 & 85.66 & 85.03 & 78.45 & 87.90  \\
\cline{3-11}
&  & Add Ham-Spam Sentence & 5044 & 7854 & 29 & 2157 & 85.51 & 84.84 & 78.45 & 87.78  \\
\hline\hline
\multirow{4}{*}{Attention}
& - & Attack Free & 5037 & 9997 & 36 & 14 & 99.67 & 99.68 & 99.86 & 99.75  \\
\cline{2-11}
& \multirow{3}{*}{Sentence-Level} 
& Add Ham Sentence & 4916 & 10000 & 157 & 11 & 98.89 & 99.12 & 99.89 & 99.17  \\
\cline{3-11}
&  & Add Spam Sentence & 5060 & 6185 & 13 & 3826 & 74.55 & 78.37 & 61.78 & 76.32  \\
\cline{3-11}
&  & Add Ham-Spam Sentence & 4916 & 6185 & 157 & 3826 & 73.59 & 76.88 & 61.78 & 75.64  \\
\hline\hline
\multirow{4}{*}{Transformer}
& - & Attack Free &  4895 & 9963 & 178 & 48 & 98.50 & 98.64 & 99.52 & 98.88  \\
\cline{2-11}
& \multirow{3}{*}{Sentence-Level} 
& Add Ham Sentence & 4696 &  9952 & 377 & 59 & 97.11 & 97.55 & 99.41 & 97.86  \\
\cline{3-11}
&  & Add Spam Sentence & 4807 & 9391 & 266 & 620 & 94.13 & 92.91 & 93.81 & 95.50  \\
\cline{3-11}
&  & Add Ham-Spam Sentence & 4696 & 9391 & 377 & 620 & 93.39 & 92.24 & 93.81 & 94.96  \\
\hline\hline
\multirow4{*}{DistilBERT}
& - & Attack Free & 5061 & 10001 & 12 & 10 & 99.85 & 99.84 & 99.90 & 99.89\\
\cline{2-11}
& \multirow{3}{*}{Sentence-Level} 
& Add Ham Sentence & 4983 & 10008 & 90 & 3 & 99.38 & 99.52 & 99.97 & 99.54 \\
\cline{3-11}
&  & Add Spam Sentence & 4896 & 10008 & 177 & 3 & 98.81 & 99.10 & 99.97 & 99.11 \\
\cline{3-11}
&  & Add Ham-Spam Sentence & 4842 & 10000 & 231 & 11 & 98.40 & 98.76 & 99.89 & 98.80 \\
\hline\hline
\end{tabular}}
\label{tab_swSentenceTrec}
\end{table*}